\documentclass[12pt]{article}   

\pdfoutput=1

\usepackage{geometry,bbm}   
\usepackage{ulem}
\usepackage[toc,page]{appendix}             		
\usepackage{graphicx,cancel}
\usepackage[usenames,dvipsnames]{color}
\usepackage{slashed}					
\usepackage{amssymb,amsmath}
\usepackage{hyperref}
\usepackage{mathtools}
\usepackage{cite}
\usepackage{subcaption} 
  \usepackage{tikzit}
  \usepackage{tikz}
  \usepackage[percent]{overpic}
\usepackage{graphicx}
\usepackage{bm}
\def\bea{\begin{eqnarray}}
\def\eea{\end{eqnarray}}
\usepackage{latexsym}
\usepackage{epsfig,amssymb,euscript}
\usepackage{amsmath}
\usepackage{array,calc,epsfig}
\usepackage{bbold}

\newcommand{\be}{\begin{equation}}
\newcommand{\ee}{\end{equation}}

\numberwithin{equation}{section}

\begin{document}

\begin{titlepage}

\begin{center}
{\LARGE
{\bf
On non-supersymmetric fixed points \vskip 17pt in five dimensions
}}
\end{center}

\bigskip
\begin{center}
{\large
Matteo Bertolini, Francesco Mignosa and Jesse van Muiden}
\end{center}

\renewcommand{\thefootnote}{\arabic{footnote}}

\begin{center}
\vspace{0.2cm}
 {\it SISSA  \\ Via Bonomea 265, I 34136 Trieste, Italy} 
\\ and \\
 {\it INFN - Sezione di Trieste \\ Via Valerio 2, I-34127 Trieste, Italy}

\vskip 5pt
{\texttt{bertmat, fmignosa, jvanmuid @sissa.it}}
\end{center}

\vskip 5pt
\noindent
\begin{center} {\bf Abstract} \end{center}
\noindent
We generalize recent results regarding the phase space of the mass deformed $E_1$ fixed point to a full class of five-dimensional superconformal field theories, known as $X_{1,N}$. As in the $E_1$ case, a phase transition occurs as a supersymmetry preserving and a supersymmetry breaking mass deformations are appropriately tuned. The order of such phase transition could not be unequivocally determined in the $E_1$ case. For $X_{1,N}$, instead, we can show that at large $N$ there exists a regime where the phase transition is second order. Our findings give supporting evidence for the existence of non-supersymmetric fixed points in five dimensions. 
\vspace{1.6 cm}
\vfill

\end{titlepage}

\newpage
\tableofcontents

%%%%%%%%%%%%%%%%%%%%%%%
%%%%%%%%%%%%%%%%%%%%%%%
\section{Introduction and summary of results}

Five-dimensional field theories, although perturbatively non-renormalizable, show interesting UV dynamics. After the pioneering work of \cite{Seiberg:1996bd,Morrison:1996xf,Intriligator:1997pq,Aharony:1997ju,Aharony:1997bh,DeWolfe:1999hj}, in recent years several interesting results have been obtained for theories with supersymmetry, using a variety of different techniques, see for instance \cite{Diaconescu:1998cn, DelZotto:2017pti, Xie:2017pfl, Jefferson:2017ahm, Jefferson:2018irk, Bhardwaj:2018yhy, Bhardwaj:2018vuu, Apruzzi:2018nre, Closset:2018bjz,Hayashi:2019jvx, Apruzzi:2019vpe, Apruzzi:2019opn, Apruzzi:2019enx, Bhardwaj:2019jtr, Bhardwaj:2019fzv, Bhardwaj:2019ngx, Saxena:2020ltf, Apruzzi:2019kgb, Closset:2019juk, Bhardwaj:2019xeg, Bhardwaj:2020gyu, Eckhard:2020jyr, Bhardwaj:2020kim, Hubner:2020uvb, Bhardwaj:2020ruf, Bhardwaj:2020avz, Closset:2020afy, Apruzzi:2021vcu, Closset:2021lhd,Collinucci:2021ofd,DeMarco:2022dgh,Legramandi:2021uds,DHoker:2016ysh,DHoker:2016ujz,DHoker:2017mds,DHoker:2017zwj,Gutperle:2017nwo,Gutperle:2018axv,Gutperle:2018vdd,Gutperle:2020rty}. 
In particular, a whole zoo of superconformal field theories (SCFT) have been shown to exist in five dimensions, some of which corresponding to possible UV completions of supersymmetric gauge theories. These conformal theories enjoy many interesting properties, such as enhancement of global symmetries \cite{Seiberg:1996bd,Morrison:1996xf, Kim:2012gu, Bergman:2013aca, Bergman:2013koa, Bergman:2013ala, Hayashi:2019jvx,Genolini:2022mpi,DelZotto:2022fnw} and emergence of non-perturbative Higgs branches \cite{Cremonesi:2015lsa, Ferlito:2017xdq, Cabrera:2018ann, Cabrera:2018jxt, Bourget:2019aer, Bourget:2019rtl, Cabrera:2019dob, Grimminger:2020dmg, Bourget:2020gzi, Bourget:2020asf, Akhond:2020vhc, Bourget:2020mez, Bourget:2020xdz, Akhond:2021knl, Akhond:2021ffo, Bao:2021ohf}. 

A still open question is whether non-supersymmetric conformal field theories exist in five dimensions. Some investigations have been carried out using $\epsilon$-expansion and bootstrap techniques, providing hints for their existence (see  
\cite{Fei:2014yja,Nakayama:2014yia,Bae:2014hia,Chester:2014gqa,Li:2016wdp,Arias-Tamargo:2020fow,Li:2020bnb,Giombi:2019upv,Giombi:2020enj, Florio:2021uoz, DeCesare:2021pfb} for some recent works). However, it is fair to say that we are still far from having a clear understanding about the existence of interacting conformal field theories (CFT) in five dimensions without supersymmetry.

A possible strategy to tackle this problem is to start from some known SCFT and deform it by a supersymmetry breaking deformation and study the corresponding RG-flow. Its end-point could be, at least in a certain region of the parameter space, a CFT. This approach was pursued in  \cite{BenettiGenolini:2019zth} where a non-supersymmetric deformation of the $E_1$ SCFT \cite{Seiberg:1996bd} was considered and conjectured to end in a CFT in the IR. In a subsequent work \cite{Bertolini:2021cew} it was shown that this deformation induces an instability in the system but that tuning the supersymmetry breaking deformation with the supersymmetry preserving deformation that the $E_1$ theory admits, a phase transition would emerge in the IR. A natural question is what the order of this phase transition actually is. 

In brane-web language the phase transition is due to an instability of the web against recombination, in analogy with what happens with systems of branes at angles in flat space, see {\it e.g.} \cite{Hashimoto:2003xz}. Neglecting interactions between the branes constituting the brane web, the energies of the two competing webs, the connected and the reconnected ones, can be easily computed and shown to dominate as a dimensionless parameter is above (resp. below) a critical value. At such critical value the two configurations are instead degenerate in energy and a phase transition between them occurs. At this level of approximation this looks first order. This is however a crude estimate since, as shown {\it e.g.} in \cite{Giveon:2007fk} in a similar context, brane interactions are important in determining the actual order of the phase transition. 

While the tree-level computation can be easily done for the $E_1$ deformed brane web, this is not so when brane interactions are taken into account. Hence, for the $E_1$ theory the order of the phase transition cannot be safely established using these techniques. What we do in this work is to consider  generalizations of the $E_1$ theory, the so-called $X_{1,N}$ SCFTs \cite{Bergman:2018hin}, which behave similarly to the former but for which, thanks to the possibility of taking $N$ large, brane interactions can be reliably computed. Interestingly, we will show that there exists a large range of parameters where the phase transition is second order and hence a non-supersymmetric CFT exists. 

The rest of the paper is organized as follows. In section \ref{Recombination}, using brane web language, we review the analysis of \cite{Bertolini:2021cew} of the $E_1$ theory and show, following the approach discussed above, that a phase transition occurs which, neglecting brane interactions, looks first order for any value of the parameters. While brane interactions are difficult to compute for this system, in section \ref{X1N} we consider a generalization of the $E_1$ theory, the $X_{1,N}$ theory, which admits a similar supersymmetry breaking deformation. In section \ref{PTX1N} we analyze this system at large $N$, in which the effect of brane interactions can be reliably computed, and show that there exists a region of the parameter space where the corresponding phase transition is in fact second order. We conclude in section \ref{discussion} with a summary of our results and a discussion on some open questions.

%%%%%%%%%%%%%%%%%%%%%
%%%%%%%%%%%%%%%%%%%%%
\section{Phase diagram of mass deformed $\mathbf{E_1}$ theory: a review}
\label{Recombination}

\begin{figure}[h]
	\centering
	\includegraphics[scale=0.34]{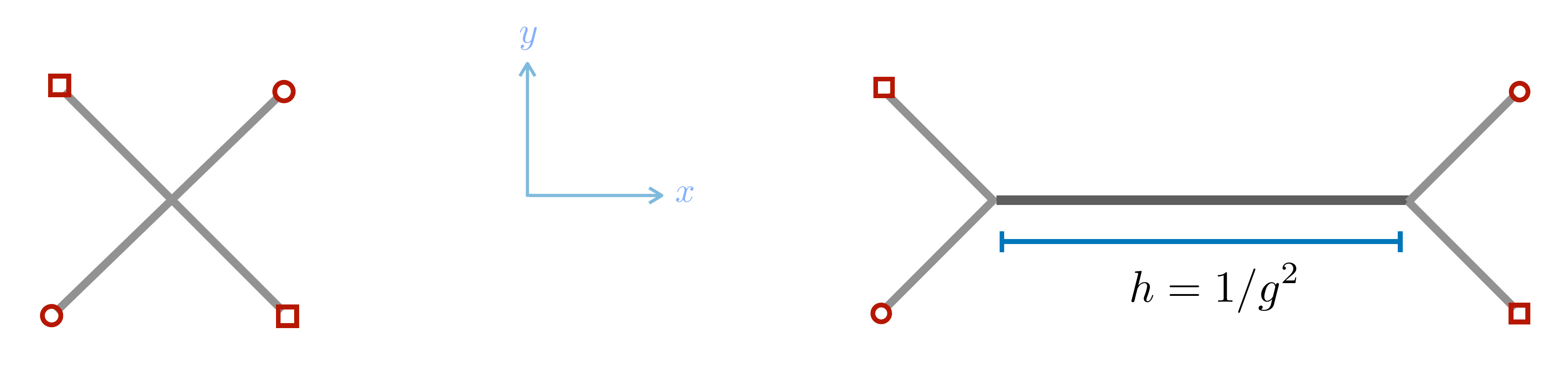}
	\caption{On the left the brane web describing the $E_1$ fixed point. On the right its supersymmetric mass deformation. Segments are (p,q) 5-branes intersecting at points on the $(x,y)$ plane. Their orientation is aligned with the corresponding (p,q) charges. Red circles/squares are [1,1] and [1,-1] 7-branes, respectively, which are orthogonal to the $(x,y)$ plane.} \label{E1SYM}
\end{figure}

The $E_1$ theory is an interacting five-dimensional SCFT which, upon a supersymmetric relevant deformation with parameter $h \equiv 1/g^2$, flows in the IR to pure $SU(2)$ SYM, with gauge coupling $g$ \cite{Seiberg:1996bd}. The corresponding brane webs, describing the interacting fixed point and $SU(2)$ SYM, respectively, are reported in figure \ref{E1SYM} (we refer to \cite{Aharony:1997ju,Aharony:1997bh,DeWolfe:1999hj}, whose conventions we adopt, for details on the use of brane webs to describe five-dimensional field theories).

\begin{figure}[h]
	\centering
	\includegraphics[scale=0.40]{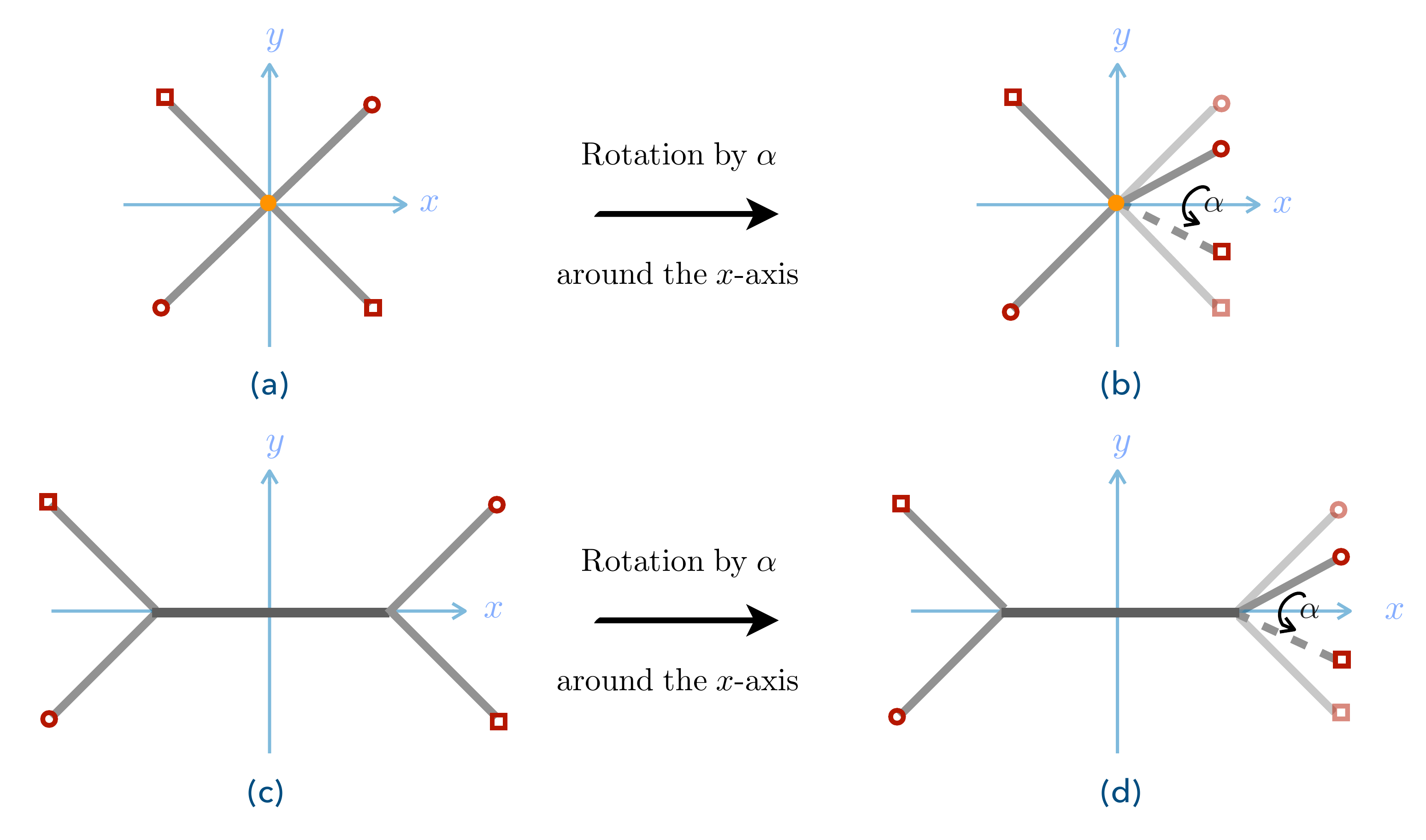}
	\caption{The supersymmetry breaking deformation of the $E_1$ theory at infinite (above) and finite (below) coupling. At infinite coupling the $(1,1)$ strings stretched between the two (half) $(1,1)$ 5-branes develop a tachyonic mode. At finite coupling the four-brane junction split and the strings get stretched and have a minimal distance of order $h$. This provides a positive mass squared contribution which competes with the tachyonic one.
	} 
	\label{E1sb}
\end{figure}

As shown in \cite{BenettiGenolini:2019zth}, the $E_1$ theory admits also a relevant  supersymmetry breaking deformation, parametrized by a mass squared  parameter $\widetilde m$. Also this deformation can be described in brane web language \cite{Bertolini:2021cew}. It corresponds to a rotation of an angle $\alpha \sim \widetilde m \, \alpha'$ of the two right 5-branes around the $x$-axis, as described in figure \ref{E1sb}.  

\begin{figure}[ht]
	\centering
	\includegraphics[scale=0.62]{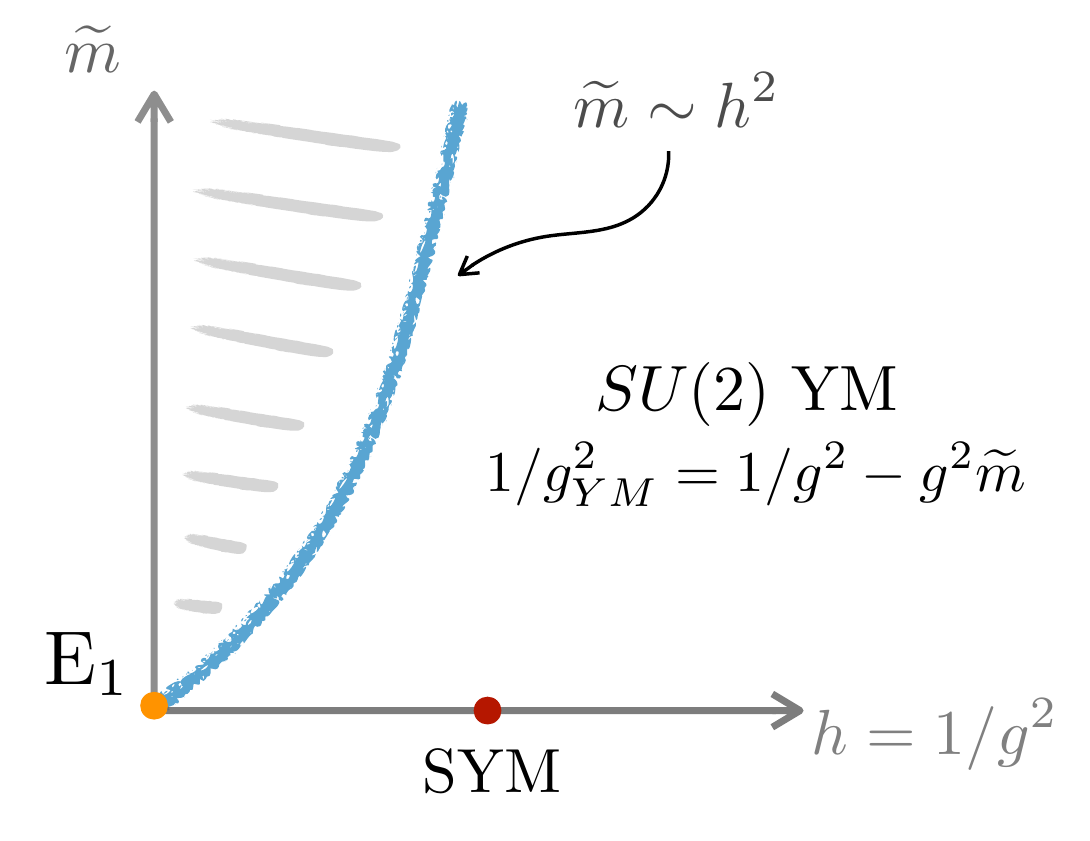}
	\caption{Phase diagram of softly broken $E_1$ theory for positive $\widetilde m$ and $h$. The white region is described by pure $SU(2)$ YM at low energy and enjoys a $U(1)_I \times U(1)_R$ global symmetry. The dashed region is a symmetry broken phase, $U(1)_I \times U(1)_R \rightarrow U(1)_D$. Along the blue line a phase transition occurs. The $SU(2)$ YM effective gauge coupling diverges there. 
	} 
	\label{E1pd}
\end{figure}

At $h=0$ the $(1,1)$ strings stretching between the $(1,1)$ 5-branes at angle, develop a tachyonic mode and the system becomes unstable towards brane recombination \cite{Hashimoto:2003xz}. At finite $h$ this same mode gets also a positive mass square contribution, since the $(1,1)$ strings gets stretched due to the opening of the four-brane junction. This contribution competes with the tachyonic one and for $h^2 > \widetilde m$ the overall mass square becomes positive. Hence,  there exist two qualitative different regions as one varies the parameters $h$ and $\widetilde m$. 
For $h^2 > \widetilde m$ the theory flows to pure $SU(2)$ YM in the IR while for $h^2 < \widetilde m$ an otherwise preserved global symmetry is spontaneously broken and the theory enters a new phase.\footnote{While for $h=0$ the tachyon potential is runaway, evidence were given in \cite{Bertolini:2021cew} that finite $h$ contributions may affect the potential and stabilize the scalar VEV at finite distance in field space.} The Yang-Mills and the symmetry broken phases are separated by a phase transition at $h^2 \sim \widetilde m$. A description of the resulting phase diagram is reported in figure \ref{E1pd}.  

An interesting aspect that was emphasized in \cite{Bertolini:2021cew} is that while the phase transition occurs at finite value of the (bare) gauge coupling $h= 1/g^2$, the $SU(2)$ (renormalized) gauge coupling $g_{\mbox{\tiny YM}}$ diverges there. So, if the phase transition were second order, the fixed point would represent a UV-completion of pure $SU(2)$ YM. The latter would emerge as the IR (free) fixed point of a RG-flow triggered by a relevant deformation of the CFT, to be identified, in the IR, with the gauge coupling, very much like what happens for the $E_1$ fixed point and ${\cal N}=1$ $SU(2)$ SYM. 

\begin{figure}[h]
	\centering
	\includegraphics[scale=0.27]{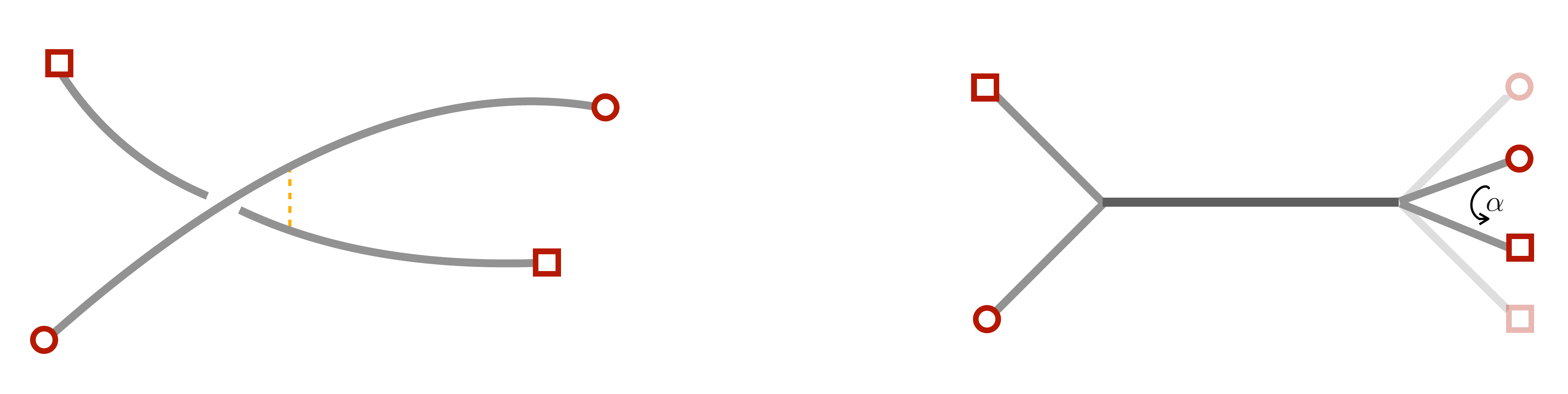}
	\caption{On the left the recombined brane web after tachyon condensation. The $(1,1)$ and $(1,-1)$ 5-branes are separated in a direction transverse to the $(x,y)$ plane. On the right the connected supersymmetry breaking configuration. Each three junction is supersymmetric but the whole system breaks supersymmetry since the boundary conditions of the D5-branes (horizontal line) on the two three-junctions are mutually non-BPS. 
	} 
	\label{Bw2}
\end{figure}

For generic values of $h$ and $\widetilde m$ there exist two brane webs compatible with charge conservation, a recombined smooth configuration after tachyon condensation and the original connected one, as described qualitatively in figure \ref{Bw2}. Following \cite{Bertolini:2021cew}, we expect the former to dominate for $h^2 < \widetilde m$ and the latter for $h^2 > \widetilde m$. At $h^2 \sim \widetilde m$ a phase transition between these topological distinct configurations is expected and one could wonder whether using brane web dynamics the order of such phase transition can be determined.

The energy of the two configurations and, in turn, the way the transition between the two brane webs occurs depends on the interaction between the constituent branes. This is hard to compute, in particular in a non-supersymmetric setup as the one we are interested in. Let us then analyze the brane system by neglecting brane interactions, first.

In this limit, the energies of the two configurations are nothing but the tensions of the various branes shaping them. In the calculation, the 7-branes on which the 5-branes end furnish a regulator, since this way the otherwise semi-infinite 5-branes become finitely extended and their energy finite.

\begin{figure}[h]
	\centering
	\includegraphics[scale=0.40]{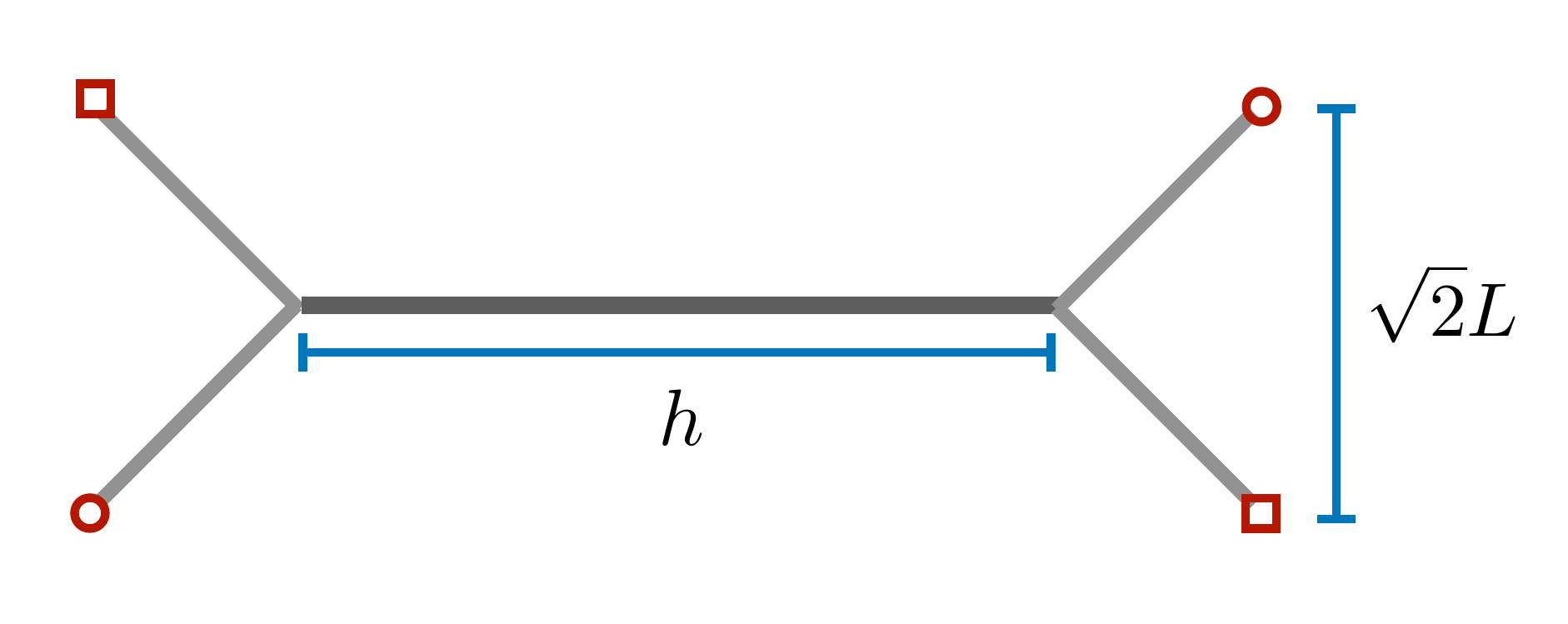}
	\caption{The connected configuration for $\alpha=0$. The (1,1) and (1,-1) 5-brane segments have all length $L$. } \label{SUSYweb}
\end{figure}

The 5-brane constituting the brane webs are of different kinds and so are their tensions. In particular, the tension of a (p,q) 5-brane is
\begin{equation}
	T_{(p,q)} = \sqrt{p^2 +q^2}~ T_{(1,0)}~,
\end{equation}
where $T_{(1,0)}$ is the D5-brane tension and, following \cite{Aharony:1997bh}, the complexified Type IIB string coupling has been set to its self-dual point, $\tau=i$.  With this in mind, let us start considering the connected configuration in the supersymmetric limit, as shown in figure \ref{SUSYweb}. The energy of this configuration  is easily computed to be 
\begin{equation}
	\label{enconalpha}
	E_{\text{con.}}(h,L)= 4\sqrt{2} L +2h
\end{equation} 
in units of $T_{(1,0)}$.
If we now rotate the right junction by an angle $\alpha \leq \pi$ around the $x$-axis as in figure \ref{E1sb}, keeping the angle between the $(1,1)$ and the $(1,-1)$ 5-brane fixed,\footnote{This can be shown to be the configuration minimizing the energy.} the energy remains the same since all lengths remain fixed. Hence, the total energy of the connected configuration in the limit in which brane interactions are neglected equals \eqref{enconalpha} for any $\alpha$. 

\begin{figure}[h]
	\centering
	\includegraphics[scale=0.50]{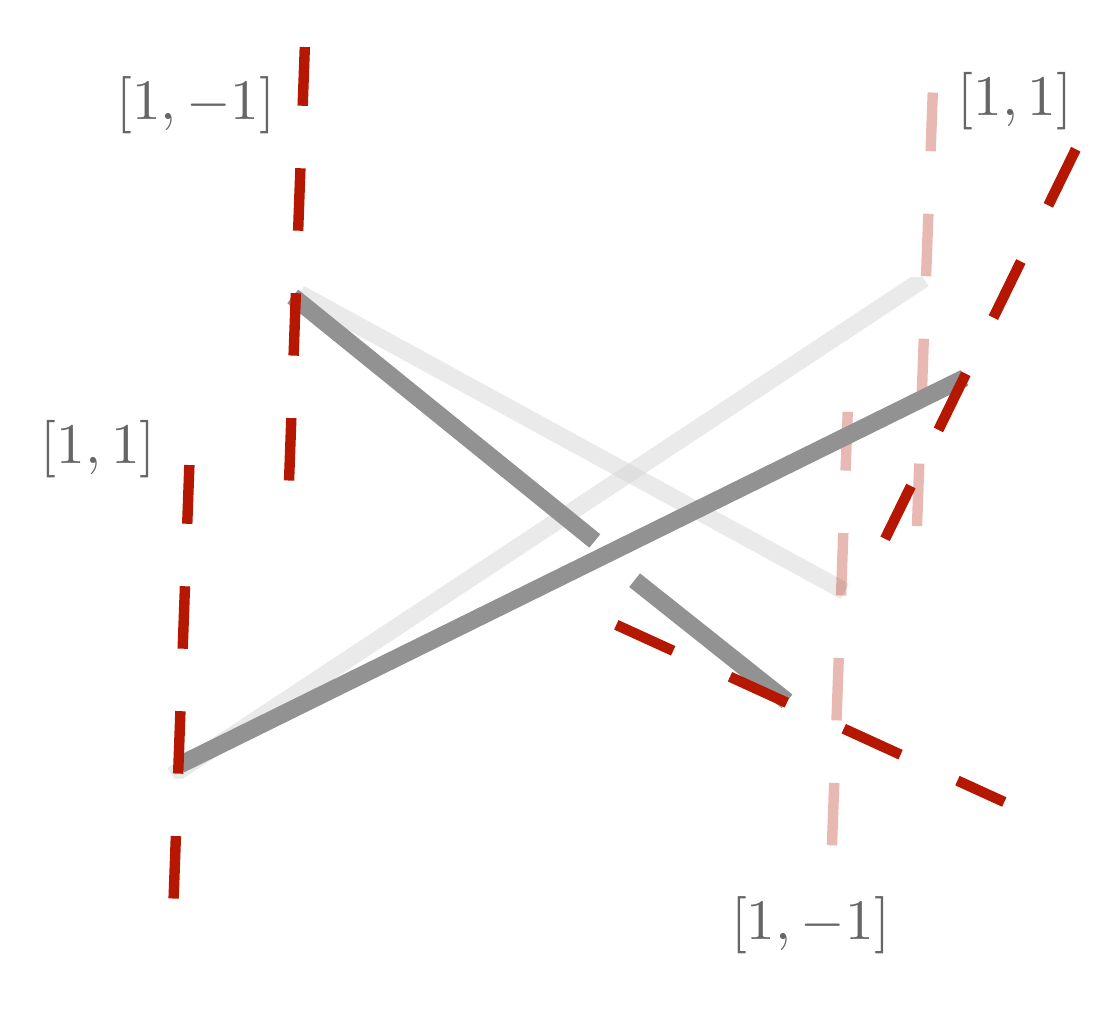}
	\caption{Recombined configuration neglecting brane interactions. Fixing the boundary conditions, that is the positions of the 7-branes, the minimal energy configurations are straight lines, as indicated.} \label{recombinedconf}
\end{figure}

In this same limit, the reconnected configuration compatible with charge conservation is nothing but the straight brane version of the brane web on the left of figure \ref{Bw2}. This comes from merging of the $(1,1)$ 5-brane prongs into a unique straight $(1,1)$ 5-brane suspended between the $[1,1]$ 7-branes, and similarly for the $(1,-1)$ 5-branes, as shown in figure \ref{recombinedconf}. 
The energy $E_{\text{rec.}}$ of the configuration depends now on the rotation angle $\alpha$ and reads
\begin{equation}
	\label{enrecalpha}
	E_{\text{rec.}}(h,L,\alpha)= 2\sqrt{2} \sqrt{(h+\sqrt{2} L)^2+ 2 L^2\cos^2\frac{\alpha}{2}}~.
\end{equation}
Comparing eqs.~\eqref{enconalpha} and \eqref{enrecalpha} we see that the connected configuration is the one with minimal energy and hence is the true vacuum of the theory for $h> h^\ast$, while the reconnected one has minimal energy for $h<h^\ast$, where 
\begin{equation}
	\label{phasetransition}
	h^\ast=\sqrt{2} L\sqrt{1-\cos\alpha}~.
\end{equation}
At $h=h^\ast$ the two configurations, which exist and remain distinct for any value of $h$, are degenerate in energy and there is a phase transition between them (in the supersymmetric limit, $\alpha=0$, the transition occurs at $h^\ast=0$ and, consistently, the connected configuration is always dominant). 
The corresponding phase diagram is similar to figure \ref{E1pd} and suggests that the phase transition, at least at this level of the analysis, is actually first order.\footnote{The same result was found independently by Oren Bergman and Diego Rodriguez-Gomez.} In particular, in both cases the transition depends on the angle $\alpha$. However, in the phase diagram of figure \ref{E1pd}, the transition point $h^\ast$ is proportional to the string length $l_s$, while in our configuration, which is completely semi-classical, there is no dependence on this parameter.\footnote{We find here a spurious dependence on the parameter $L$ that we used to regulate. We will further comment about it when considering the effects of brane interactions, in section \ref{PTX1N}.}

One might wonder if anything could change once brane interactions are taken into account. In fact, it is expected brane interactions to affect the order of the phase transition, as it was shown to be the case in {\it e.g.} \cite{Giveon:2007fk}, where four-dimensional gauge theories were studied using rather similar  brane models. One of the key ingredients of the analysis of \cite{Giveon:2007fk} was the possibility of selecting a regime where few constituent branes could be studied as probes in the background of many others, and take advantage of the gravitational background generated by the latter. This is something we cannot achieve in our case, since our brane web is composed by one $(1,1)$, one $(1,-1)$ and, once $h \not =0$, two $(1,0)$ 5-branes and none of them can be treated as a probe in the background of the others. So, in order to take advantage of an approach as in \cite{Giveon:2007fk} a generalization of the $E_1$ theory is required. A natural such candidate is the so-called $X_{1,N}$ theory \cite{Bergman:2018hin}, whose structure will be reviewed in the next section.

%%%%%%%%%%%%%%%%%%%%%%%%%%%%%
%%%%%%%%%%%%%%%%%%%%%%%%%%%%%
\section{Generalizations of $\mathbf{E_1}$: the $\mathbf{X_{1,N}}$ theory}
\label{X1N}

The brane web of $N$ $(1,1)$ branes intersecting $M$ $(1,-1)$ branes at a 90 degrees angle realizes the so-called $X_{M,N}$ fixed point \cite{Bergman:2018hin}. 
\begin{figure}[h]
	\centering
	\includegraphics[scale=0.30]{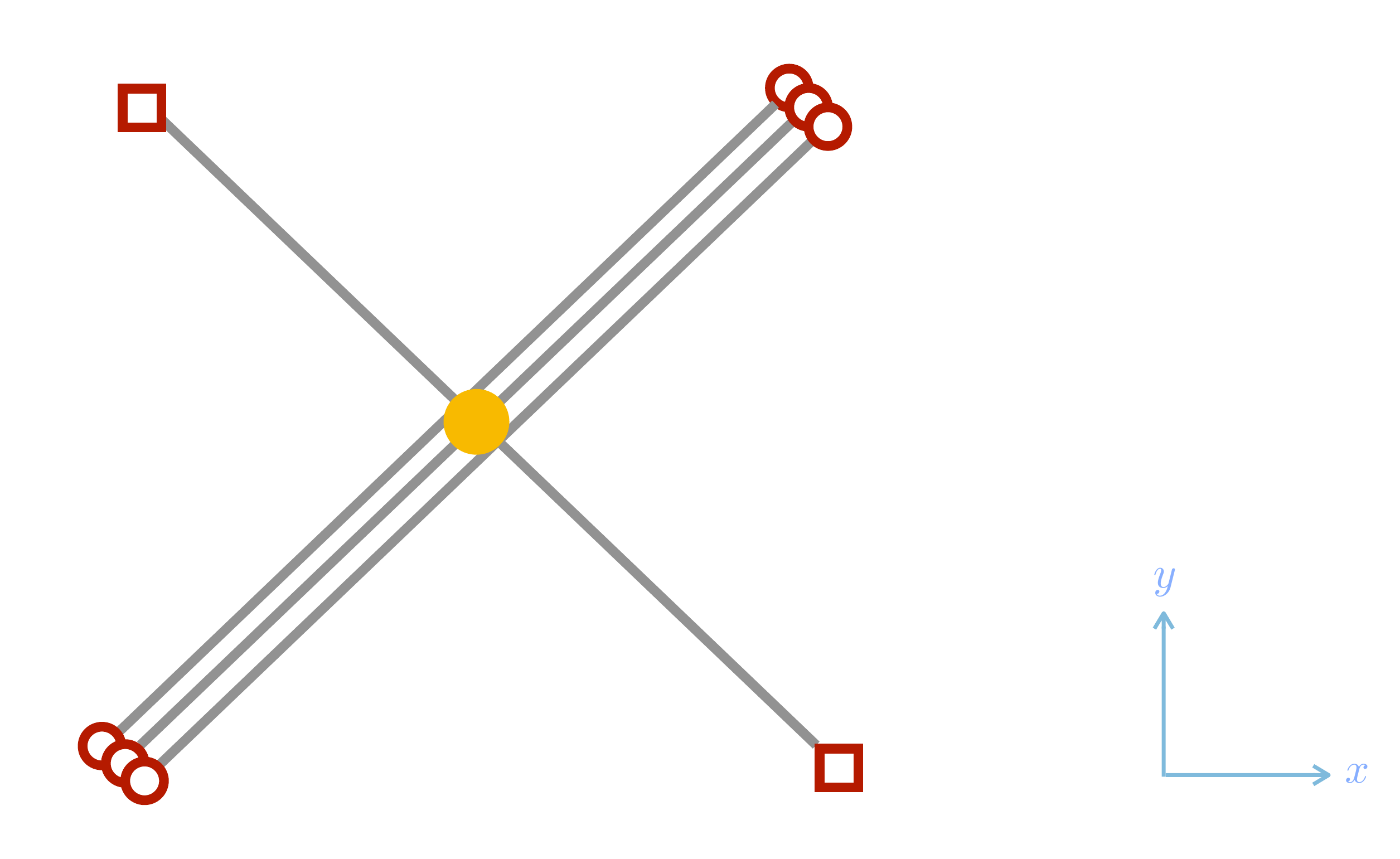}
	\caption{$X_{1,N}$ fixed point ($N=3$ in the figure).} 
	\label{pqwebN}
\end{figure}
Specializing to the case $M=1$, the web reduces to the one in figure \ref{pqwebN}. 

\begin{figure}[h]
	\centering
	\includegraphics[scale=0.20]{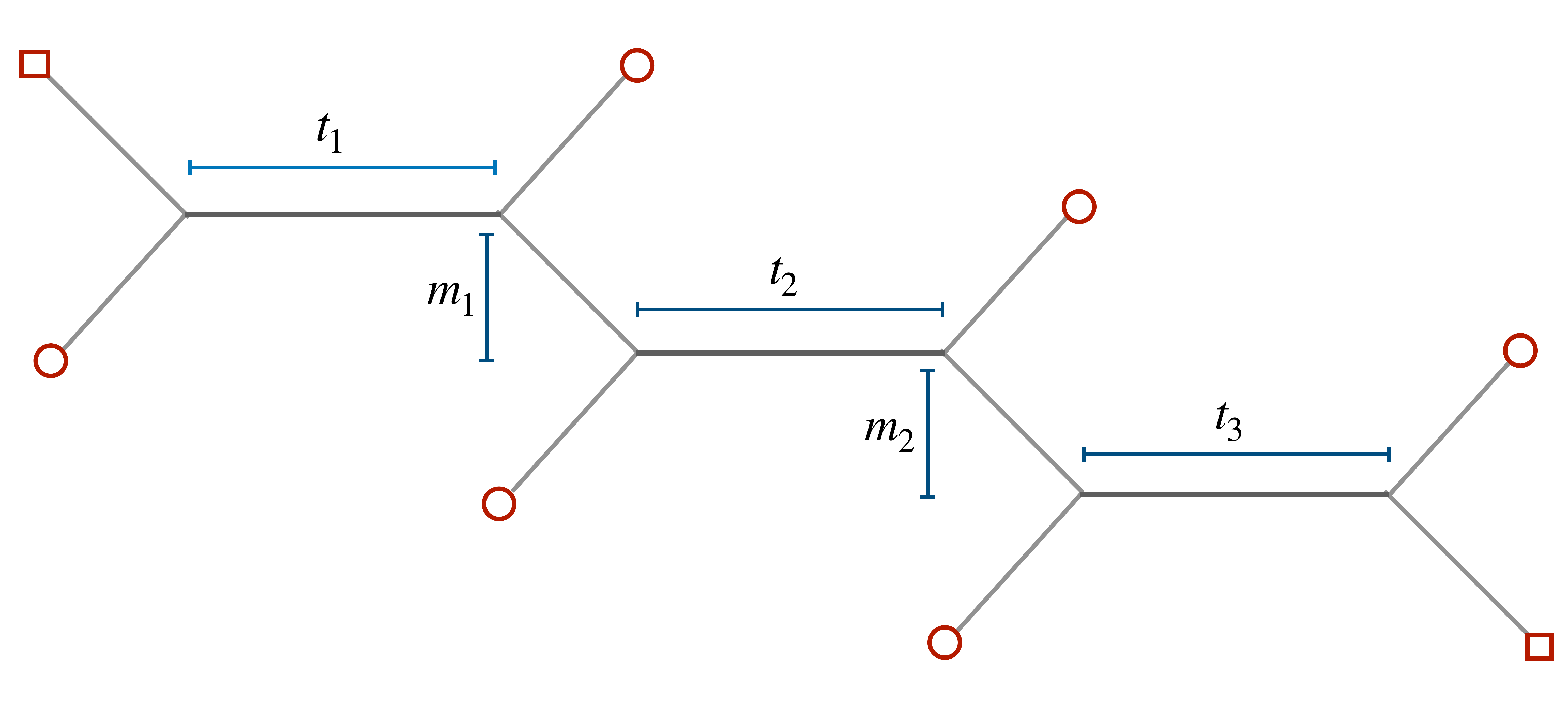}
	\caption{The $X_{1,N}$  brane web in the $SU(2)^N$ limit.} \label{WeakCouplingN}
\end{figure}

\begin{figure}[h]
	\centering
	\includegraphics[scale=0.17, trim={0, 12cm, 0, 12cm}, clip]{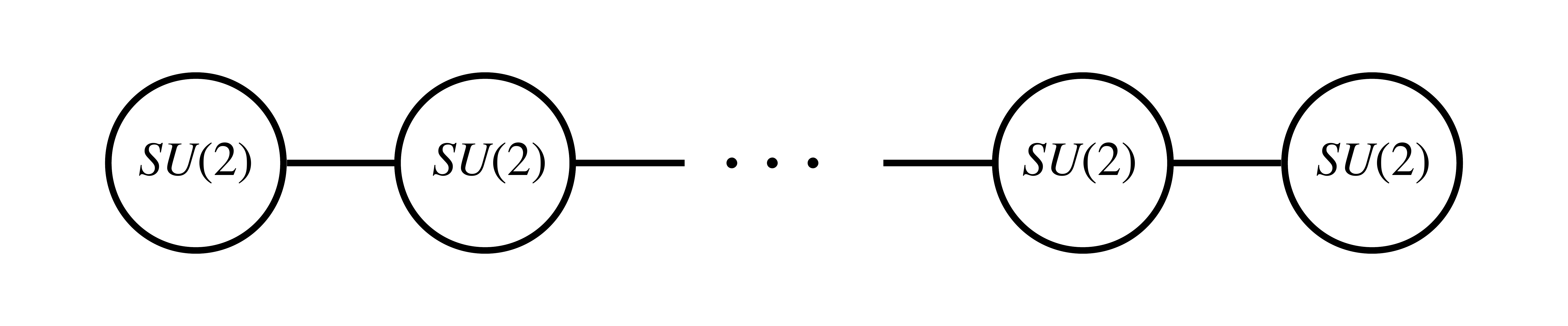}
	\caption{$SU(2)^N$ quiver.} \label{SU2Quiver}
\end{figure}

Similarly to the $E_1$ theory, one can switch on (the now several) supersymmetric relevant deformations. These trigger an RG-flow and drive the theory to a supersymmetric gauge theory in the IR. This corresponds to opening-up the brane web as shown in figure \ref{WeakCouplingN}, while figure \ref{SU2Quiver} is the quiver diagram describing such low energy effective theory. 
This is a $SU(2)^N$ supersymmetric gauge theory with matter in the bifundamental.  

The lengths of the $(1,0)$ branes, that we dub $t_i$ in the following, correspond to the square of the inverse (effective) gauge couplings of the $SU(2)$ gauge factors. The vertical distance between the D5-branes associated with the $i-$th and the $(i+1)-$th groups defines instead the mass $m_i$ of the corresponding bifundamental. 

For generic values of $t_i$ and $m_i$ the global symmetry of the system is $U(1)_{I}^N \times U(1)_{F}^{N-1}$. Similarly to what happens for $E_1$, when $t_i=0$ the instantonic $U(1)_{I}$ associated with the $i-$th node enhances to $SU(2)_{I}$. This is manifest from the brane web: when $t_i=0$ one can make two 7-branes of the same type (either $[1,1]$ or $[1,-1]$) to lie on top of each other, hence enhancing the 8-dimensional gauge symmetry living on their world-volume, which corresponds to an instantonic symmetry in the five-dimensional theory \cite{DeWolfe:1999hj,Bertolini:2021cew}. Similarly, when $m_i=0$, two $(1,1)$ 5-branes become aligned, inducing an enhancement of the corresponding flavor symmetry from $U(1)_{F}$ to $SU(2)_{F}$.

At the fixed point, the global symmetry is believed to get enhanced to $SU(2N)$ \cite{Bergman:2013aca}. This can be understood from the brane web by the possibility of superimposing the $2N$ $[1,1]$ 7-branes at the fixed point, see figure \ref{pqwebN}.\footnote{Strictly speaking, this argument is a bit naive since no affine extension of the $A_{N-1}$ algebra can be constructed from systems of 7-branes \cite{DeWolfe:1998pr}. So the standard methods used in presence of exceptional symmetries \cite{DeWolfe:1999hj} cannot be applied.} This also implies that the Higgs branch, parametrized at weak coupling by the massless bifundamentals, gets enhanced. At the fixed point, this is the $2N$-dimensional minimal nilpotent orbit $\overline{\mathcal{O}_{[2^N]}(\mathfrak{su}(2N))}$ , as can be shown by drawing the corresponding magnetic quiver. This is nothing but the space of $2N\times 2N$ complex matrices $M$ with $M^2=0, \text{Tr}M=0$ or the Higgs branch of four-dimensional $U(N)$ supersymmetric gauge theory with $2N$ flavors.\footnote{We thank Antoine Bourget for elucidating this point to us.} 

In the following we will consider a supersymmetry breaking deformation of the $X_{1,N}$ theory very similar to the one we discussed previously for the $E_1$ theory. Again, the existence of a phase transition in the space of parameters will be manifest. However,  very much like what was done in \cite{Giveon:2007fk}, in this case the possibility to play with the large $N$ limit will let us get some insights on the nature of this phase transition. In particular, we will show that in a certain range of parameters the phase transition is actually second order, and a non-supersymmetric fixed point is then expected to exist in the phase diagram.

%%%%%%%%%%%%%%%%%%%%%%%%
%%%%%%%%%%%%%%%%%%%%%%%%
\section{Phase transitions in the $\mathbf{X_{1,N}}$ theory}
\label{PTX1N}

\begin{figure}[h]
	\centering
	\includegraphics[scale=0.28]{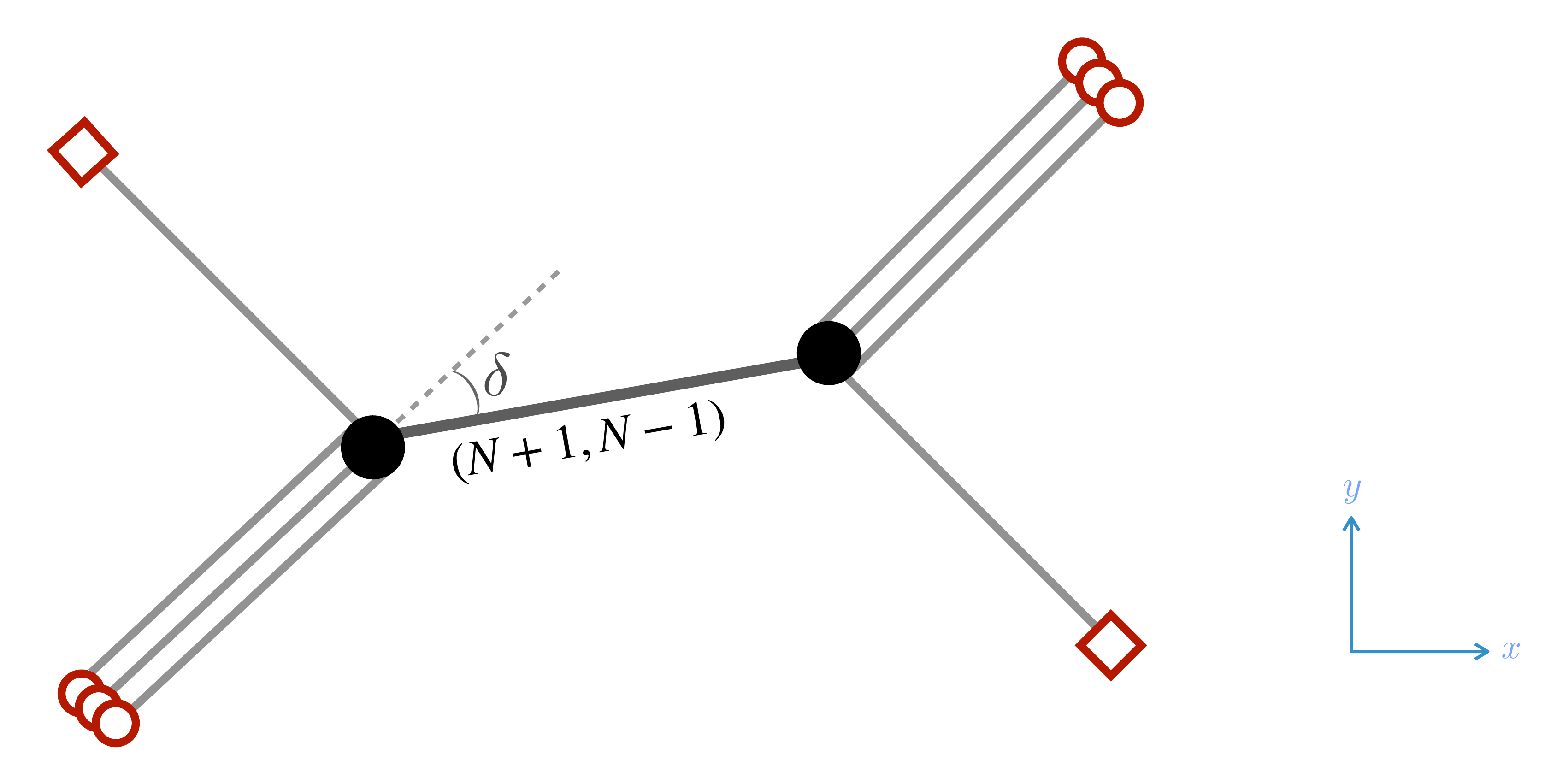}
	\caption{Opening the fixed point via a supersymmetric deformation with parameters $t_i=-2m_i= h$. The $(1,-1)$ 5-branes remain at a $90$ degrees angle with the $(1,1)$ 5-brane stack, while the larger $N$ the smaller the angle $\delta$ between the stack and the $(N+1,N-1)$ 5-brane of length $h$.} 
	\label{pqwebNMassive1}
\end{figure}

Let us consider a deformation of the $X_{1,N}$ theory with parameters $t_i=-2m_i= h$ for all $i$. This makes the single junction of the fixed point theory to  separate  into two, as shown in figure \ref{pqwebNMassive1}: the $(1,1)$ 5-branes remain perpendicular to the $(1,-1)$ ones while $(N+1,N-1)$ represents the intermediate (p,q) 5-brane, whose length equals $h$. The $SU(2N)$ flavor symmetry is broken to   $SU(N)_L\times SU(N)_R\times U(1)_B$ while the $SU(2)$ R-symmetry remains unbroken, since the deformation takes place in the $(x,y)$ plane, only.

It is worth noting that, for generic $N$, this deformation does not give any simple five-dimensional field theory, but rather a limit in which some of the gauge couplings of the $N$ $SU(2)$ gauge factors diverge\footnote{For instance, for $N=2$ after the deformation $t_1=t_2= - 2m$, we get a theory where both gauge couplings of the two $SU(2)$ nodes diverge.}. An exception is the case $N=1$ for which the mass deformation leads to pure $SU(2)$ SYM.

Exactly as we did for $E_1$, we can now break supersymmetry by rotating the right brane junction by an angle $\alpha$ around the axis along which the $(N+1,N-1)$ 5-brane is aligned. The deformation involves the transverse directions to the $(x,y)$ plane and hence affects now also the $SU(2)$ R-symmetry, which gets broken to its Cartan. This has a natural field theory counterpart. The supersymmetry preserving deformation corresponds to give a non-vanishing VEV to the lowest component of the background vector multiplet associated with the global symmetry current, which is a singlet under the $SU(2)$ R-symmetry. Here, instead, we give a VEV to a highest component which, as such, breaks supersymmetry. This is a triplet under $SU(2)$ and so the R-symmetry is broken to $U(1)$, very much like what happens for the $E_1$ theory \cite{BenettiGenolini:2019zth,Bertolini:2021cew}. 

\begin{figure}[h]
	\centering
	\includegraphics[scale=0.25]{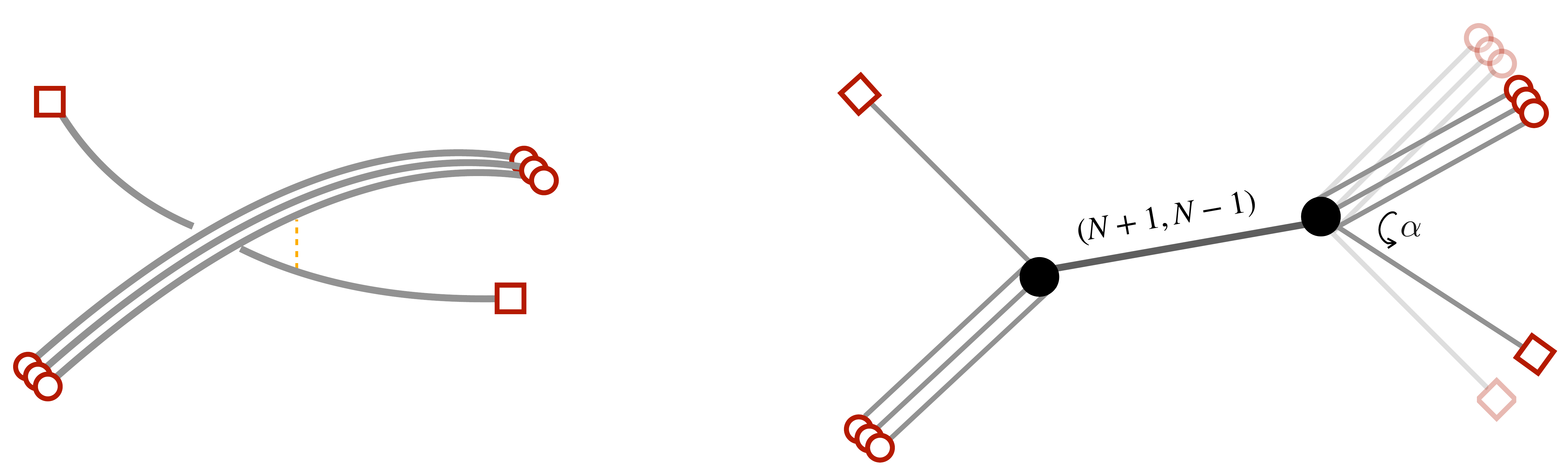}
	\caption{The two competing brane webs after the supersymmetry breaking deformation. The recombined system consists of one $(1,-1)$ 5-brane and $N$ $(1,1)$ 5-branes, separated in a direction transverse to the $(x,y)$ plane.}
	\label{Bw2x}
\end{figure}

From the structure of the brane web, and comparing with figure \ref{Bw2}, one could argue the effects of the supersymmetry breaking deformation to be qualitatively similar to what happens for the $E_1$ theory \cite{Bertolini:2021cew}. A scalar mode is expected to become tachyonic for small enough $h$ and the brane web wants to recombine. The two competing configurations, compatible with brane charge conservation, are shown in figure \ref{Bw2x}. Their energies, in the limit in which brane interactions are neglected, are a generalization of eqs.~\eqref{enconalpha}-\eqref{enrecalpha} and read 
\begin{equation}\label{Nenconrec}
	\begin{aligned}
		E_{\text{rec.}} =&\, \sqrt{2}\,[f(\sin\delta)+N f(\cos\delta)] ~~,~~f(a) \equiv \sqrt{(h+2L a)^2+4L^2(1-a^2)\cos^2\frac{\alpha}{2}}\,, \\
		E_{\text{con.}} =&\,\sqrt{2}\left[2 N L+2L+\sqrt{N^2+1}\,h\right]~,
	\end{aligned}
\end{equation}
where the reconnected configuration is the natural generalization to $N>1$ of the straight brane configuration of figure \ref{recombinedconf}. 
It is possible to show that also in this case there exists a (single) critical value $h^\ast$ that separates two regions in the space of parameters where one brane web dominates against the other, and viceversa. 

\begin{figure}[h]
	\centering
	\includegraphics[scale=0.25]{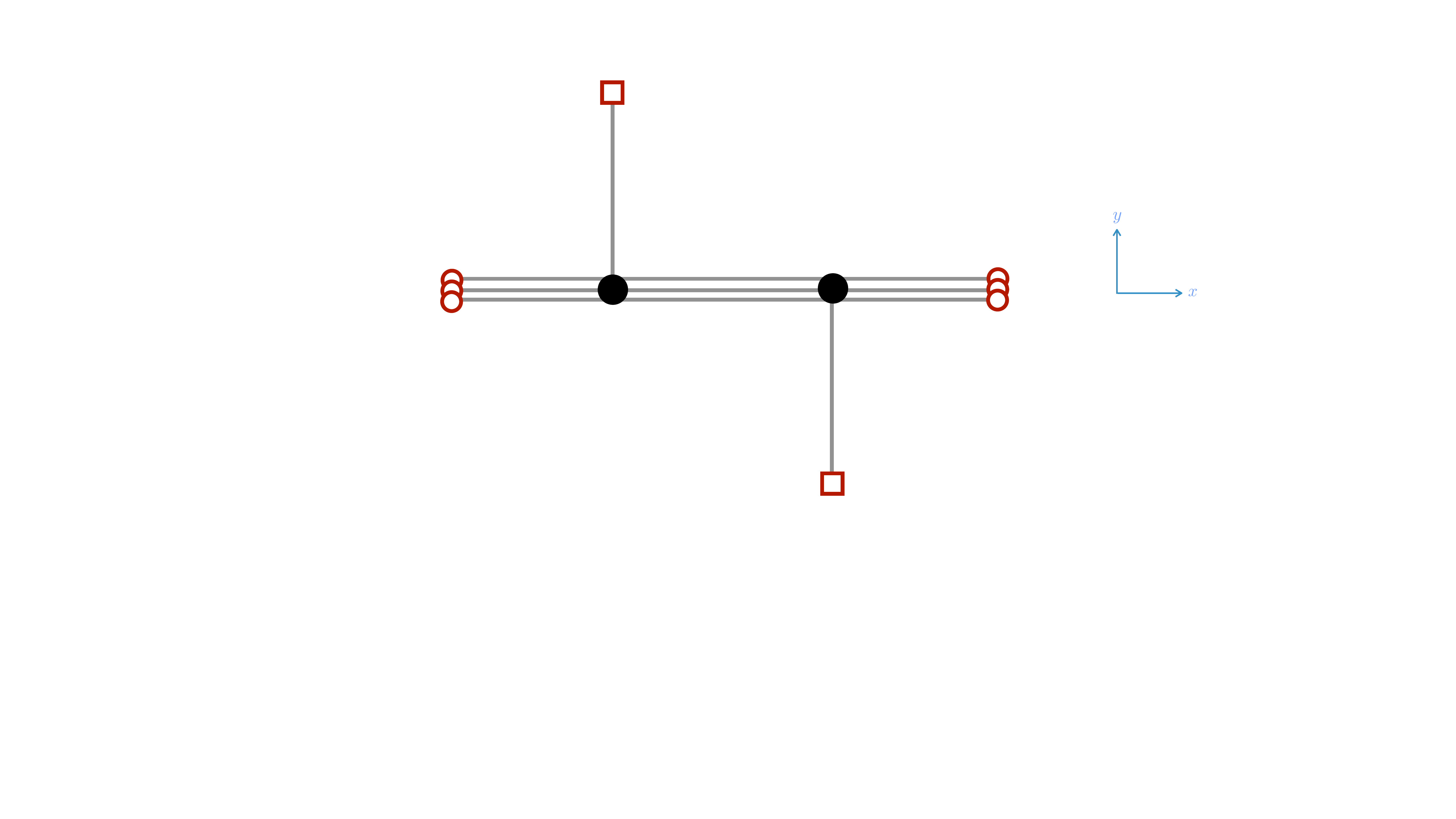}
	\caption{The deformed $X_{1,N}$ theory in the large $N$ limit. The system becomes that of $N$ $(1,1)$ 5-branes on which two $(1,-1)$ 5-branes ends.} 
	\label{NpqwebNMassive1}
\end{figure}

So far this is no different from what we discussed in section \ref{Recombination}, and neglecting interactions the phase transition looks again first order. The point, now,  is that we can consider $N$ to be parametrically large. This has two effects. The first is that it makes easier to compute brane interactions in the recombined brane system, left of figure \ref{Bw2x}, since in the large $N$ limit this can be treated as a probe $(1,-1)$ 5-brane in the gravitational background of $N$ $(1,1)$ 5-branes. The second  effect is that it makes the angle $\delta$ between the two stacks of $N$ $(1,1)$ 5-branes and the $(N+1,N-1)$ 5-brane going to zero
\be
\cos \delta = \frac{N}{\sqrt{N^2+1}} ~~,~~ \lim_{N \rightarrow \infty} \delta = 0~,
\ee
while the $(N+1,N-1)$ 5-brane becomes indistinguishable from a stack of $N$ $(1,1)$ 5-branes. Hence, in the strict $N \to \infty$ limit the system in figure \ref{pqwebNMassive1} reduces to that in figure \ref{NpqwebNMassive1}. In this limit brane charge conservation at brane junctions does not force the $N$ stack to bend anymore (and to change its nature) due to the $(1,-1)$ branes which end on it. The energies (\ref{Nenconrec}) of the two configurations simplify as
\begin{equation}
	\begin{aligned}
		E_{\text{rec.}} =&\, \sqrt{2}\,\left[N(h+2L) +\sqrt{h^2+4L^2 \cos^2\frac{\alpha}{2}}\,\,\right]+\mathcal{O}\left(1/N\right) \,, \\
		E_{\text{con.}} =&\, \sqrt{2}\left[N(h+2L)+2L\right] +\mathcal{O}\left(1/N\right)~, 
	\end{aligned}
\end{equation}
with the transition point being at $h^\ast= 2L\sin \alpha/2$. 

We note, in passing, that in this limit our system becomes very similar to the one considered in \cite{Giveon:2007fk}, yet in one dimension higher. This will be useful later.

%%%%%%%%%%%%%%
\subsection{Phase transitions in the backreacted $\mathbf{X_{1,N}}$ brane-web}
\label{PTN}

In this section we will take brane interactions into account and see how the nature of the phase transition discussed previously may change. 

\begin{figure}[h]
	\centering
	\includegraphics[scale=0.27]{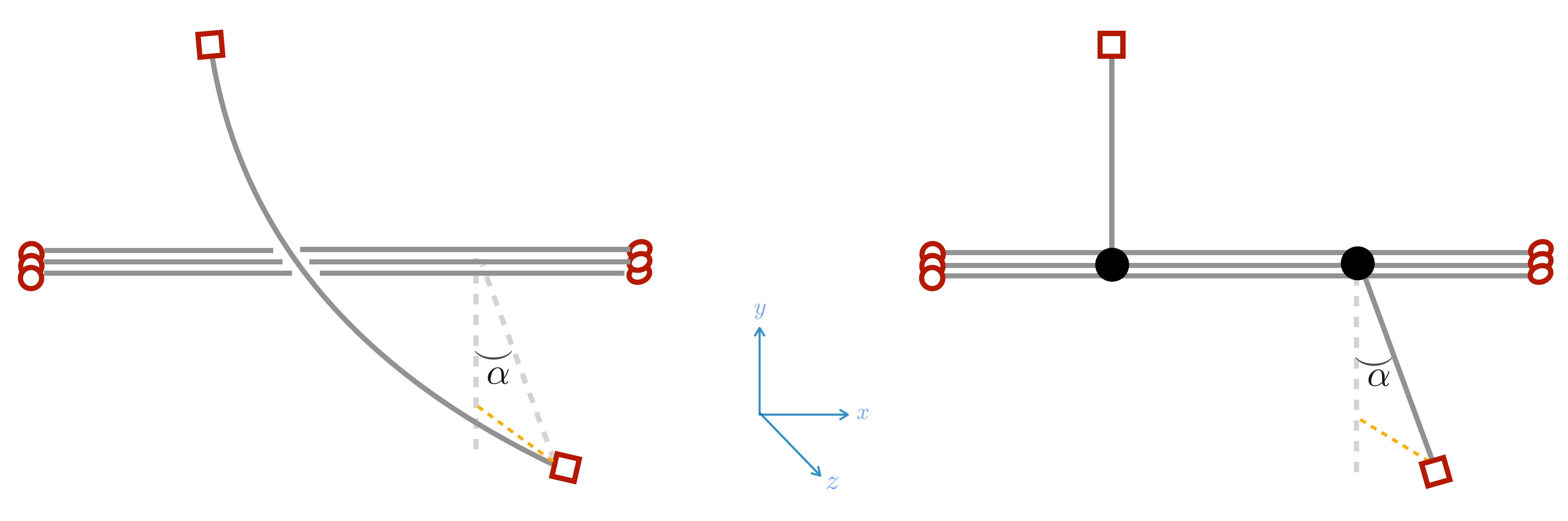}
	\caption{The two competing brane webs after the supersymmetry breaking deformation, in the large $N$ limit.}
	\label{Bw3x}
\end{figure}

As already noticed, in the large $N$ limit the original supersymmetric configuration simplifies to the one depicted in figure \ref{NpqwebNMassive1}. Rotating by an angle $\alpha$ the non-supersymmetric connected and reconnected brane webs look instead as in figure \ref{Bw3x}. 

The difference in energy between the two configurations depends on the $(1,-1)$ 5-brane only, since the $(1,1)$ 5-brane stack is unperturbed in this limit, as in the non-interacting case. Hence, its contribution will be factored out in what follows, and we will just compute the $(1,-1)$ 5-branes energy. The system can be treated as a probe $(1,-1)$ 5-brane in the gravitational background of $N$ $(1,1)$ 5-branes which can exert a force on (and hence bend) the probe brane. Note, however, that by the very geometry of the problem, this does not happen for the connected brane web, right of figure \ref{Bw3x}, whose energy is then the same as when interactions are neglected, eq.~\eqref{Nenconrec}. In what follows, we will hence compute the effects of brane interactions on the left brane web of figure \ref{Bw3x}.

Let us start considering the background generated by the $(1,1)$ branes stack. We can align these branes along the $01234x$ directions, while the $(1,-1)$ branes, in the supersymmetric configuration, are aligned along $01234y$. It is useful to introduce cylindrical coordinates  as
\begin{equation}\label{Eq: cylindrical coords}
	(x,y,z) = (x, \rho\cos\phi,\rho \sin\phi)\,.
\end{equation}
In these coordinates the $N$ $(1,1)$ 5-branes are located along $(x,0,0)$ while the $(1,-1)$ 5-brane, after the supersymmetry breaking deformation, has boundary conditions on the $[1,-1]$ 7-branes it ends on $P_1\equiv(x_1, \rho_1, 0)$ and $P_2\equiv(x_2, \rho_2 \cos\varphi, \rho_2 \sin\varphi)$,  where 
\be
\label{phialpha}
\varphi = \pi - \alpha~. 
\ee
The supersymmetric limit corresponds to $\varphi=\pi$. In these cylindrical coordinates the back-reacted metric of the $N$ $(1,1)$ 5-branes takes the form\footnote{We present the metric in Einstein frame, and take asymptotic values of the axio-dilaton equal to $\tau_0 = i$, as before.}
\begin{equation}
\label{metric1}
	ds^2_{10} = H^{-1/4} ds_{\mathbb{R}_{1,4}}^2 + H^{-1/4} dx^2 +H^{3/4} (d\rho^2 + \rho^2 d\phi^2 + ds^2_{\mathbb{R}_2})  \,, \quad H = 1 + \frac{\ell^2}{\rho^2}\,,
\end{equation}
where $\ell= 2^{1/4} l_s \sqrt{N}$. To simplify notations we will measure quantities in units of $\ell$, and re\"instate the correct factors through dimensional analysis when needed. The axio-dilaton equals
\begin{equation}
	\tau = C_0 + i \,e^{-\Phi} = \frac{1+H}{1-H} + 2\, i \frac{\sqrt{H}}{1+H}\,,
\end{equation}
while the three forms have support on the $S^3$ sphere surrounding the stack
\begin{equation}
\frac{1}{(2\pi l_s)^2}\int_{S^3} F_3= N, \,\,\,\,\,\, \frac{1}{(2\pi l_s)^2}\int_{S^3} H_3= N.
\end{equation}
The brane action of the $(1,-1)$ 5-brane consists of a DBI term and a WZW term, given by 
\begin{equation}
	S_{(1,-1)} =- T_{(1,0)} \int d^6 \xi\, \Delta(\tau,\bar \tau)\sqrt{-\det\left( P[g_{\mu\nu}] + \frac{P[C_2 - B_2]}{\Delta(\tau,\bar \tau)}  \right)}+ T_{(1,0)} \int \left(C_6 -B_6\right)~.
\end{equation}
where
\begin{equation}
	\Delta(\tau,\bar \tau) = \sqrt{\frac{2 i (1-\tau)(1-\bar{\tau})}{\tau-\bar{\tau}}}~. 
\end{equation}
Note that the two-form gauge potentials are transverse to the $(1,-1)$ 5-branes, and the six-form gauge potentials are equal, thus the brane action will only depend on the ten-dimensional metric and axio-dilaton.

Filling in the metric pull-back and the value of $\tau$ one finds that
\begin{equation}\label{Lag(1,-1)}
	S_{(1,-1)} = - \sqrt{2}T_{(1,0)} \int d x \sqrt{H^{-1} + \dot\rho^2 + \rho^2 \dot\phi^2}~,
\end{equation}
where we have denoted derivatives with respect to $x$ with a dot. Modulo an overall normalization, the DBI (\ref{Lag(1,-1)}) is the same as for a D5 brane in the background of a NS5 brane \cite{Giveon:2007fk}. Indeed, the two configurations are $SL(2,\mathbb{R})$ dual. Since the corresponding Lagrangian, $\mathcal L_{(1,-1)}$, does not explicitly depend on $x$ and $\phi$ we find the following two constants of motion 
\begin{align}
	\mathcal I =& H\mathcal{L}_{(1,-1)}\,, 
	\label{Eq: constant of motion 1}\\\ 
	\mathcal Q =& H \rho^2 \dot \phi~. 
	\label{Eq: constant of motion 2}
\end{align}
We search for brane configurations ending on the points $P_1$ and $P_2$. These develop a minimum $x_m$ at which $\dot\rho(x_m)=0$. This represents the turning point of the solution, namely the minimal distance of the probe from the brane stack. Taking $\rho_m = \rho(x_m)$, we find that
\begin{equation}\label{Eq: Constant of motion turning point}
	\sqrt{1+\rho_m^2 + \mathcal Q^2 } =\rho_m \mathcal I \,.
\end{equation}
To solve the full equations of motion we split the solution into two branches $x \in [x_i,x_m]$, where $i=1,2$ and, as mentioned below  eq.~\eqref{Eq: cylindrical coords}, $x_i$ labels the positions of the $[1,-1]$ 7-branes along the $x$ direction. We can then use eqs.~\eqref{Eq: constant of motion 2} and \eqref{Eq: Constant of motion turning point} to solve eq.~\eqref{Eq: constant of motion 1} through separation of variables
\begin{equation}
\begin{aligned}
		\sqrt{1+\mathcal Q^2} (x_m - x_1) =&\,  \int\limits_{\rho_m}^{\rho_1} d \rho \frac{H}{\sqrt{H_m - H}} = \rho_m \sqrt{\rho_1^2 - \rho_m^2} + \theta_1~, \\
		\sqrt{1+\mathcal Q^2} (x_2 - x_m) =&\,  \int\limits_{\rho_m}^{\rho_2} d \rho \frac{H}{\sqrt{H_m - H}} = \rho_m \sqrt{\rho_2^2 - \rho_m^2} + \theta_2~, 
\end{aligned}
\end{equation}
where, for later convenience we have defined  $\theta_i = \arccos (\rho_m /\rho_i)$. Similarly we find, using the equation of motion for $\rho$, that
\begin{equation}
\begin{aligned}
		\sqrt{1+\mathcal Q^2}\phi_m =\,  -\mathcal Q \,\theta_1~,\quad \sqrt{1+\mathcal Q^2}(\phi_m - \varphi) =\, \mathcal Q \,\theta_2~,
\end{aligned}
\end{equation}
To further analyse the system we will assume the simplification $\rho_1 = \rho_2 \equiv L$, and thus $\theta_1 = \theta_2 \equiv \theta$, such that
\begin{equation}\label{Eq: Delta x and varphi}
	\begin{aligned}
	\sqrt{1+\mathcal Q^2} h =\, L^2 \sin2\theta + 2\theta\,,\quad 
	\sqrt{1+\mathcal Q^2}  \varphi =\, 2 \mathcal Q \,\theta\,,
	\end{aligned}
\end{equation}
where $h \equiv x_2 - x_1$. 
Solving the second equation in \eqref{Eq: Delta x and varphi} for $\mathcal Q$ we can rewrite the first equation as (re-instating the appropriate factors of $\ell$ defined below eq.~\eqref{metric1})
\begin{equation}\label{Eq: Delta x 2}
	\ell h (\theta) = \sqrt{1 - \left(\frac{\varphi}{2\theta}\right)^2}(L^2 \sin2\theta + 2 \ell^2 \theta)~,
\end{equation}
which is transcendental and does not have a closed-form expression when solving for $\theta$. Since the constant of motion $\mathcal Q$ is real, and $0 \leq \varphi \leq \pi $ we conclude that $ \varphi\leq2\theta \leq \pi$. In the supersymmetric limit this equation trivializes and one has a solution only for $h=0$. This is consistent since in this regime the reconnected and the connected brane webs become the same, while for $h \not= 0$ the reconnected one does not exist.

The energies for the reconnected and connected configurations can be now easily computed as (minus) their evaluated brane actions and read
\begin{equation}
\begin{aligned}
	\ell E_{\text{rec.}} =\, 2 \sqrt{2}T_{(1,0)} \sqrt{H_m- \left(\frac{\varphi}{2\theta}\right)^2} \rho_m   L \sin\theta ~,~ 
	E_{\text{con.}} = 2 \sqrt{2}T_{(1,0)} L~,
\end{aligned} 
\end{equation}
where, as already noticed, the energy of the connected configuration is unaffected by brane interactions and hence equals that in eq.~\eqref{Nenconrec}. This implies that 
\begin{equation}
	\left( \frac{E_{\text{rec.}}}{E_{\text{con.}}} \right)^2 = \left(1- \frac{\varphi^2}{4\theta^2 H_m}\right) \left(1+\frac{\rho_m^2}{\ell^2}\right)\left( 1- \frac{\rho_m^2}{L^2} \right)~.
\end{equation}
The natural variables of interest are the distance between the two $[1,-1]$ 7-branes along the $x$ direction $h$, the relative rotation between them $\varphi$, and their distance from the $(1,1)$ 5-branes stack $L$. To rewrite the ratio of energies in terms of these physical variables one must solve eq.~\eqref{Eq: Delta x 2} to find $\theta(h , \varphi, L)$. This requires a combination of analytical and numerical methods and will be dealt with below.   

We will first focus on the case $\alpha = \pi$, that can be studied almost completely analytically. This will be important when we move on studying the system for general values of $\alpha$, which will turn out to be qualitatively similar, albeit one must resort to numerical methods.

%%%%%%%%%%%%%%
\subsubsection*{The $\alpha = \pi$ case}
\label{Sec: alpha=pi}

Taking $\alpha = \pi$, the brane setup is $SL(2,\mathbb{R})$ dual to a D5-NS5 system that is T-dual to the D4-NS5 brane system studied in \cite{Giveon:2007fk}. Following a completely analog analysis as in \cite{Giveon:2007fk} we will give strong evidence that, in a certain range of parameters, the $X_{1,N}$ brane-web undergoes a second order phase transition. Even though the computation is cognate to the one in \cite{Giveon:2007fk} we will go through it in detail since it will provide a good intuition for the physics when $\alpha\neq \pi$.
 
Taking $\alpha=\pi$ several quantities simplify. The transcendental eq.~\eqref{Eq: Delta x 2} now becomes Kepler's equation\footnote{This equation can actually be solved analytically for $\theta$, in terms of a series of Bessel functions, within the range $-\ell \pi < h < \ell \pi$.}
\begin{equation}\label{Eq: Delta x for alpha=pi}
	\ell h(\theta) = L^2 \sin2\theta + 2 \ell^2 \theta\,,
\end{equation}
where $0\leq \theta \leq \pi/2$. A maximum for $h$ is reached at
\begin{equation}\label{Eq: max of h alpha = pi}
	\ell h_0 = L^2 \sin2\theta_0 + 2 \ell^2 \theta_0\,,\quad \text{with} \quad	L^2\cos2\theta_0 = - \ell^2\,,
\end{equation}
which can only be solved when $L \geq \ell$. In the following, we will split the analysis into two cases, $L\leq \ell$, and $L> \ell$, which will turn out being qualitatively different.
\begin{figure}
\centering
	\includegraphics[scale=0.27]{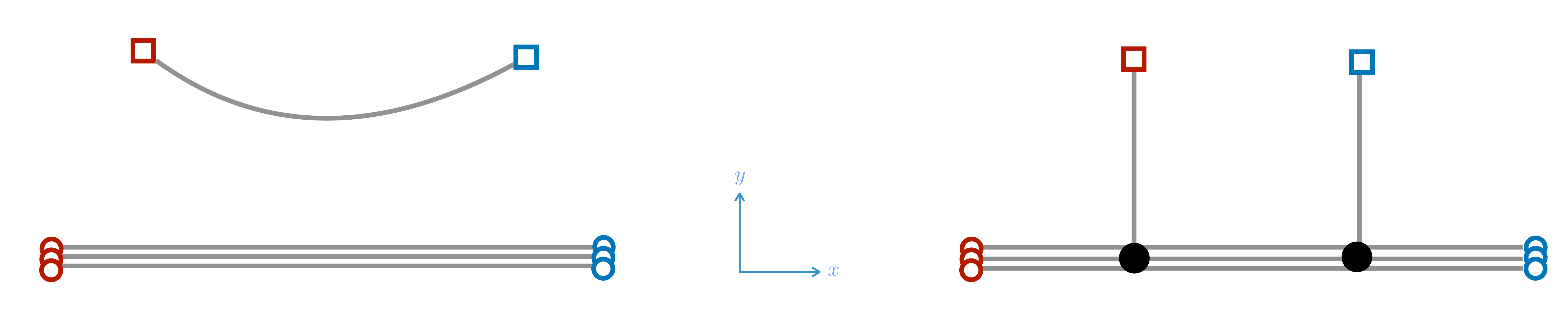}
	\caption{The reconnected and connected brane webs for $\alpha =\pi$. In this case everything happens on the $(x,y)$ plane only. Blue squares and circles refer to 7-branes orthogonal to the $(x,y)$ plane which look however as anti-branes compared to unrotated ones.} \label{Fig: brane configurations for alpha=pi}
\end{figure}

\begin{itemize}
\item $L\leq \ell$: We find that $h$ monotonically increases from $h(0) = 0$ to $h(\pi/2) = \pi \ell$. In the regime $0\leq h \leq \pi \ell$ there are thus two solutions to the brane action, the reconnected and the connected ones, whose brane webs are depicted in figure \ref{Fig: brane configurations for alpha=pi}. The ratio of their respective energies is given by 
\begin{equation}\label{Eq: Ratio of energies}
	\left( \frac{E_{\text{rec.}}}{E_{\text{con.}}} \right)^2 = \left(1+\frac{\rho_m^2}{\ell^2}\right)\left( 1- \frac{\rho_m^2}{L^2}\right) < 1\,.
\end{equation}
This ratio is always smaller than one, so we find that the energetically favorable configuration is the reconnected one. At $h  = \pi \ell$ we find that $\rho_m = 0$, the ratio goes to one and, consistently, the reconnected and the connected brane webs become degenerate. For $h>\pi \ell$ eq.~\eqref{Eq: Delta x for alpha=pi} ceases to have a solution, and thus only the connected configuration solves the equations of motion. 

Schematically we depict the distinct phases of the brane configurations through a potential in figure \ref{Fig: Potential for rho<l}. 
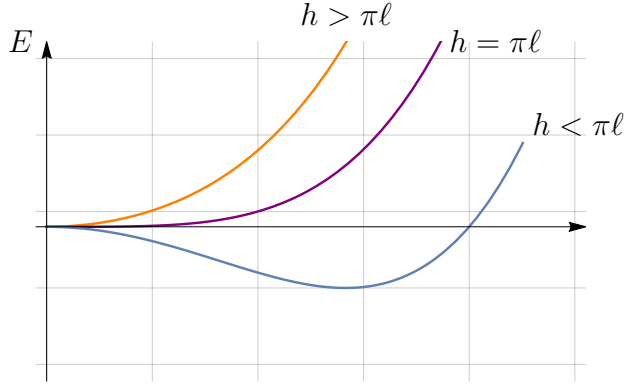
\begin{figure}[h]
	\centering
	\vspace{1.2cm}
	\begin{overpic}[scale=0.8]{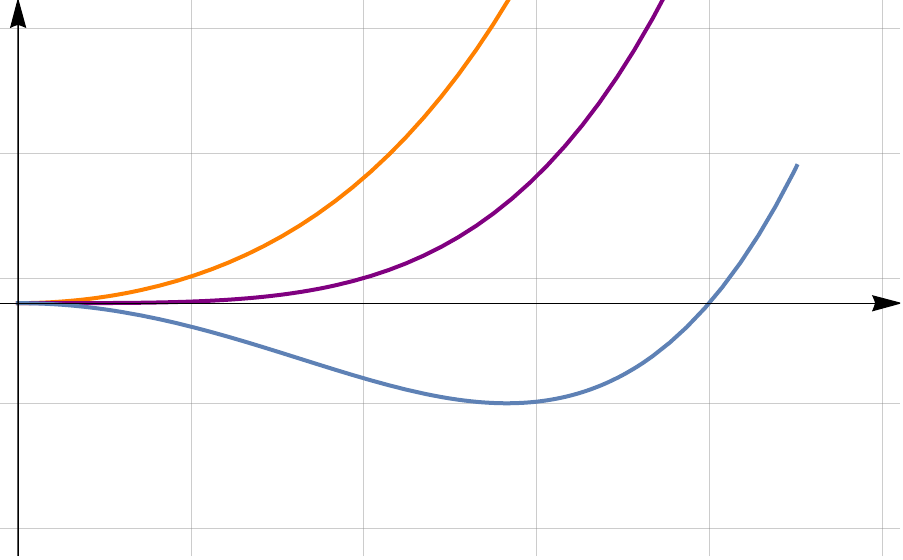}
		\put (-5,60) {$E$}
		\put (48,65) {$h > \pi\ell $}
		\put (75,60) {$h = \pi\ell$}
		\put (90,45) {$h < \pi \ell$}
	\end{overpic}
	\caption{The potential energy as a function of the configuration space of the web as $h$ is varied, for $L\leq \ell$.}\label{Fig: Potential for rho<l}
\end{figure}
Whenever $h<\pi \ell$, the potential has a minimum coinciding with the reconnected configuration and a maximum coinciding with the connected one. As the value $h$ increases, the minimum of the potential does as well, until $h = \pi \ell$, at which point the two extrema merge and the potential has a single minimum corresponding to the connected configuration. We thus find that the system undergoes a second order phase transition when $h$ passes the value  $\pi \ell = \pi \, 2^{1/4} \ell_s \sqrt{N}$.  We note, for future purpose, that this value is independent of $L$.

\item
$L > \ell$: the function $h(\theta)$ has a maximum, $h_0$, given by eq.~\eqref{Eq: max of h alpha = pi}. This maximum decreases whenever $L$ does, until $L= \ell$, at which it is at $\theta_0 = \pi/2$. Whenever $h>h_0$ there is no solution to eq.~\eqref{Eq: Delta x for alpha=pi}, and therefore only the connected configuration exists. Instead, in the region 
\begin{equation}
	h_0 \geq h \geq \pi \ell ~, 
\end{equation}
Kepler's equation has two solutions labeled by $\theta_S, \theta_L$, denoting the previous angles as, respectively, the smallest and the largest ones associated with the same value of $h$. These solutions are associated with two distinct reconnected 5-brane configurations.

For $h < \pi \ell$ one can show that
\begin{equation}\label{Eq: ratio energies 2}
	\left( \frac{E_{\text{rec.}}}{E_{\text{con.}}} \right)^2 = \left(1+\frac{\rho_m^2}{\ell^2}\right)\left( 1- \frac{\rho_m^2}{L^2}\right) = \left(1 + \frac{L^2 \cos^2 \theta}{\ell^2}\right) \sin^2 \theta \leq1\,.
\end{equation}
The reason is that the ratio is monotonically increasing in $\theta$ and smaller than or equal to $1$ for 
\begin{equation}
\label{theta*1}
	\theta \leq \theta^* \,,\quad \text{where} \quad \theta^* = \arcsin \ell/L< \theta_0\,,
\end{equation}
where $h^* = h(\theta^*)> \pi\ell$. Hence for $h < \pi \ell$ the reconnected brane configuration is always energetically favorable with respect to the connected one.

When $h>\pi\ell$ the analysis is slightly more involved. There are now three brane configurations whose energies $(E_{\text{con.}},E^S_{\text{rec.}},E^L_{\text{rec.}})$ we have to compare, where the energies $E^{S}_{\text{rec.}}, \, E^L_{\text{rec.}}$ are associated with the smooth solutions with $\theta_S, \,\, \theta_L$ respectively. Since $\pi/2>\theta_L>\theta_0$, and the ratio of energies decreases in this region, we have that
\begin{equation}
	\left( \frac{E_{\text{rec.}}^L}{E_{\text{con.}}} \right)^2 > \left( \frac{E_{\text{rec.}}(\pi/2)}{E_{\text{con.}}}  \right)^2= 1\,,
\end{equation}
with $E_{\text{rec.}}(\pi/2)$ represents the energy of the reconnected configuration with $\theta=\pi/2$. This tells us that the connected configuration is always energetically favorable compared to the reconnected one with $\theta = \theta_L$. Moreover, it can be shown that $E_{\text{rec.}}^L>E_{\text{rec.}}^S$, using the fact that the sum and differences of $\theta_L$ and $\theta_S$ are bounded by
\begin{equation}
	0\leq\theta_L + \theta_S \leq \pi\,,\quad \text{and} \quad 0\leq\theta_L - \theta_S \leq \pi/2\,,
\end{equation}
and that $h(\theta_L) = h(\theta_S)$. The discussion above shows that $E_{\text{rec.}}^S/E_{\text{con.}}$ can be either bigger or smaller than $1$, depending on the value of $h(\theta_S)$. We denote with $h^* = h(\theta^*)$ the value of $h(\theta_S)$ for which $E_{\text{rec.}}^S/E_{\text{con.}} = 1$. Schematically the different phases are depicted through a potential in Figure \ref{Fig: Potential for rho>l}. 

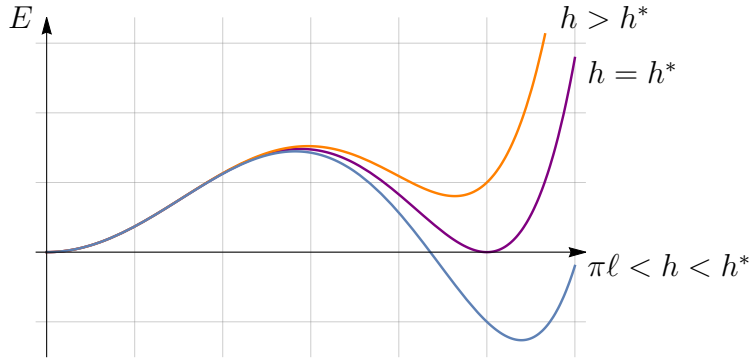
\begin{figure}[h]
	\centering
	\begin{overpic}[scale=0.8]{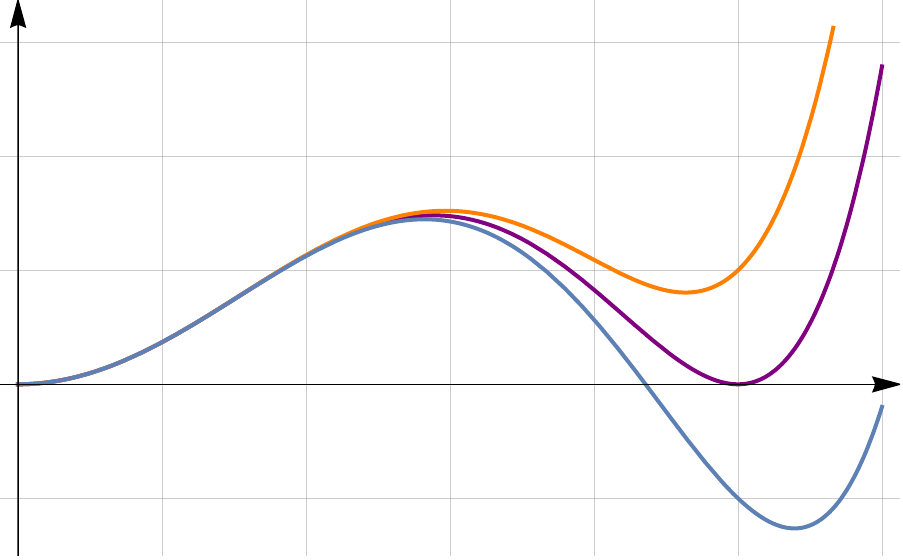}
		\put (-5,60) {$E$}
		\put (95,60) {$h > h^* $}
		\put (100,50) {$h = h^*$}
		\put (100,15) {$\pi \ell < h < h^*$}
	\end{overpic}
	\caption{The potential energy as a function of the configuration space of the web for some values of $h$, for $L> \ell$.}\label{Fig: Potential for rho>l}
\end{figure}
The connected configurations correspond to the left minimum of the potentials, the smooth reconnected solutions with $\theta = \theta_L$ correspond to the maxima of the potentials, and the smooth reconnected solutions associated with $\theta_S$ correspond to the right minima. Depending on $h$, these minima can be either local or global, showing that the brane configuration undergoes a first order phase transition when $h$ passes through $h^*$. Note that contrary to the case $L \leq \ell$, the point at which the phase transition occurs, $h^\ast$, now depends on $L$ through eq.~\eqref{theta*1}. 

\end{itemize}

%%%%%%%%%%%%%%
\subsubsection*{Generic values of $\alpha$}
\label{Sec: alpha<pi}

We now want to generalize the previous analysis to generic values of $\alpha$.  The transcendental equation is now
\begin{equation}\label{Eq: trascendentalgen}
	\ell h = \sqrt{1 - \left(\frac{\varphi}{2\theta}\right)^2}(L^2 \sin2\theta + 2 \ell^2\theta)\,,
\end{equation}
and $h$ has an extremum at 
\begin{equation}
\label{Eq: trascendentalgen2}
	L\cos 2\theta \left[ 2\theta (4 \theta^2 - \varphi^2) +  \varphi^2 \tan 2\theta \right] = - 8 \ell \theta ^3\,.
\end{equation}
Eq.~\eqref{Eq: trascendentalgen2} is not solvable analytically, so we will have to resort to numerical analysis. In this way, one can show that this equation has a zero only for 
\begin{equation}\label{Eq: rho* general alpha}
	L\geq \ell_{\varphi} = \frac{\pi \ell }{\sqrt{\pi^2 - \varphi^2}} \geq 1~,
\end{equation}
where $\ell_{\varphi}$ plays the same role as $\ell$ of previous section ($\ell_{\varphi=0}=\ell$).
\begin{figure}
	\centering
		\begin{overpic}[scale=0.65]{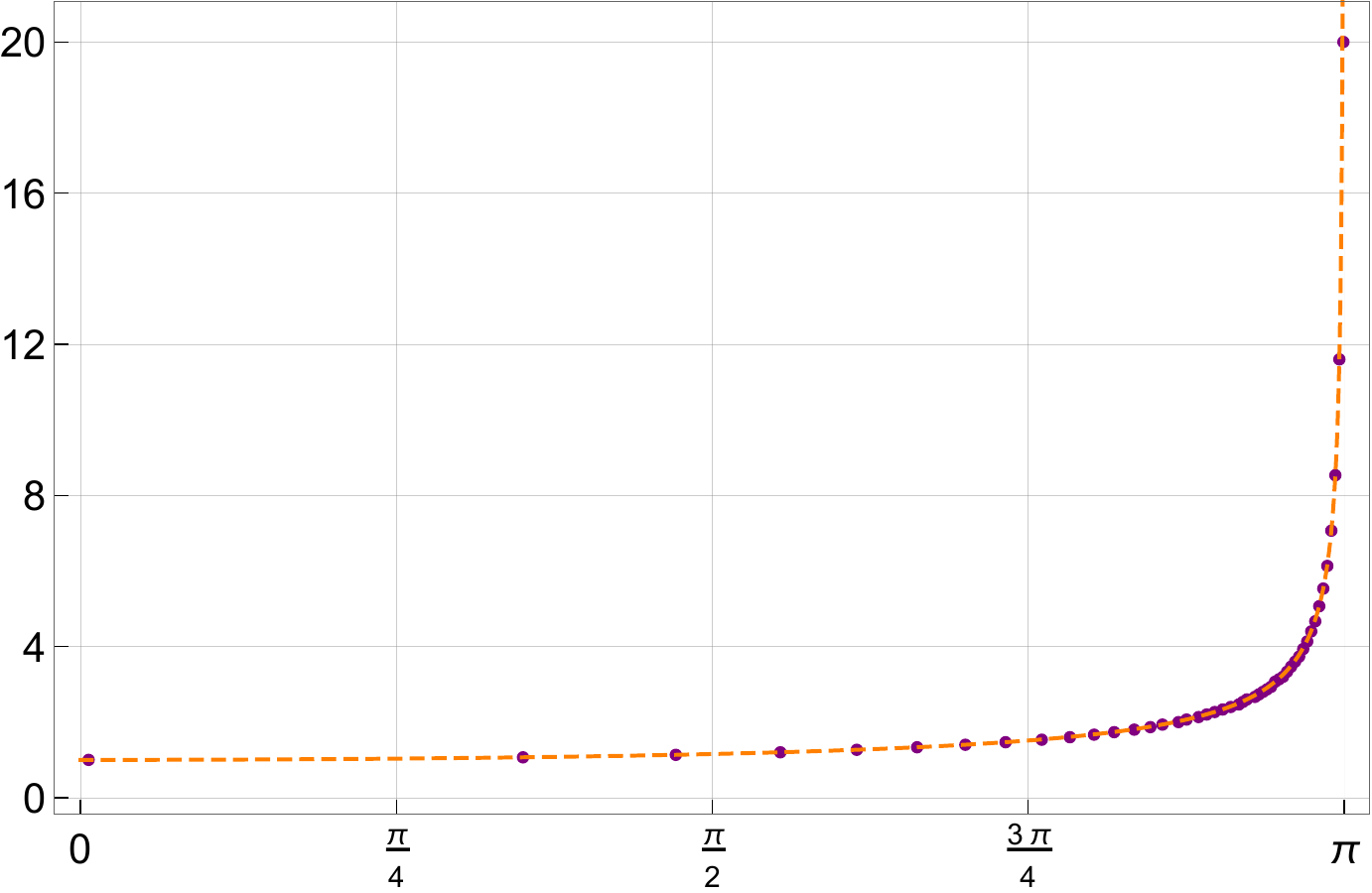}
		\put (-10,64) {$\ell_\varphi/\ell$}
		\put (101,-1) {$\varphi$}
	\end{overpic}
\vspace{0.15cm}
	\caption{Plot of $\ell_\varphi/\ell$ as a function of $\varphi$. The yellow dotted line is represents the analytical function $\pi/\sqrt{\pi^2-\varphi^2}$, and the purple dots show the numerical results.}\label{Fig: rho* of Delta phi}
\end{figure}
The function $h$ has at most one extremum, which is a maximum, when $L\geq  \ell_\varphi$. This follows from the fact that for $\pi/2 > \theta > \theta_0$, where $\theta_0$ is the value for which $h$ reaches its maximum $h_0$, the second derivative of $h$ with respect to $\theta$ is strictly negative. 

Qualitatively, $h$ behaves similarly to the case $\alpha = \pi$, just replacing $\ell \rightarrow \ell_\varphi$. In the following, we then distinguish the case $L\leq  \ell_\varphi$ from the case $L >  \ell_\varphi$. 

\begin{itemize}
\item $L\leq \ell_\varphi$: There are two brane configurations, a connected configuration and a reconnected one. Additionally, since the reconnected energy is monotonically increasing in $h$ and $h$ itself is monotonically increasing in $L$, we find that
\begin{equation}
	\left( \frac{E_{\text{rec.}}}{E_{\text{con.}}} \right)^2 \leq  \left[ 1 + \frac{\theta^2 - 4 \varphi^2}{(\pi/2)^2 - 4 \varphi^2} \frac{(\pi/2)^2}{\theta^2} \cos^2\theta \right] \sin^2\theta \leq 1\,.
\end{equation}
The ratio only saturates the bound at $\theta = \pi/2$. Therefore, when $L \leq \ell_\varphi$ the reconnected configuration is energetically favorable. When $h$ increases and crosses the value $\tilde h = \ell\sqrt{\pi^2 - \varphi^2}$, a second order phase transition occurs, after which only the connected brane configuration remains. We thus find a behavior that is qualitative the same as in the case $\alpha=\pi$.

The minimal distance $\rho_m$ between the recombined $(1,-1)$ brane and the stack decreases continuously from $\rho_m = L \cos \varphi/2$ at $h=0$ down to $\rho_m=0$ at the transition $h = \tilde h$. In the process, the reconnected brane comes closer and closer to the stack and flattens along the direction of the latter, until $\rho_m$ reaches zero. At this point, the reconnected configuration becomes indistinguishable from the connected one, as it can be shown taking the $\rho_m\rightarrow 0$ limit in the equations of motion \eqref{Eq: constant of motion 1}-\eqref{Eq: constant of motion 2}, realizing the second order phase transition. 

\item $L >  \ell_\varphi$: the function $h$ does have a maximum $h_0$, and when
\begin{equation}
	h_0 \geq h \geq \tilde h \,,\quad \text{with} \quad \tilde h = \ell \sqrt{\pi^2 -\varphi^2} \,,
\end{equation}
there exist two reconnected configurations, together with the connected one. The two reconnected configurations are again associated with two values $\theta_S \leq \theta_L$, for which $h(\theta_S) = h(\theta_L)$. Analogously to the $\alpha=\pi$ case, we denote the energies of the three configurations as $E_{\text{con.}}$, $E_{\text{rec.}}^S$, and $E_{\text{rec.}}^L$. Numerically, it is possible to show that
\begin{equation}
	\left( \frac{E_{\text{rec.}}^L}{E_{\text{rec.}}^S}  \right)\geq 1\,,\qquad \left( \frac{E_{\text{rec.}}^L}{E_{\text{con.}}}  \right)\geq 1 \,,
\end{equation}
and that $E_{\text{rec.}}^S/E_{\text{con.}}$ can be either bigger or smaller than $1$, depending if $h(\theta_S)$ is above or below a critical value $h^*$. In figure \ref{Fig: ratios 2} we show the generic behavior of the ratio of energies in function of $h$, here specifically at values $L/\ell = 2$, and $\varphi = \pi/16$, illustrating the behavior mentioned above.

Whenever $h < \tilde h$, one can argue, in a similar way as we did in the $L \leq \ell_\varphi$ case, that there is only one reconnected configuration, and that its energy is always favored over the connected one. Therefore we can conclude that if $L >  \ell_\varphi$, the brane system undergoes a first order phase transition when $h$ increases and crosses a value $h^*$, as in the $\alpha=\pi$ case.
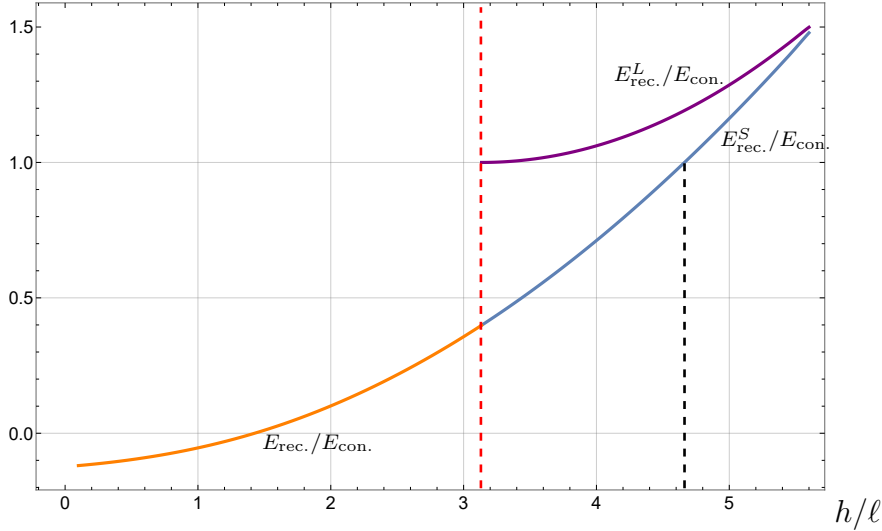
\begin{figure}[h]
	\centering
	\begin{overpic}[scale=0.48]{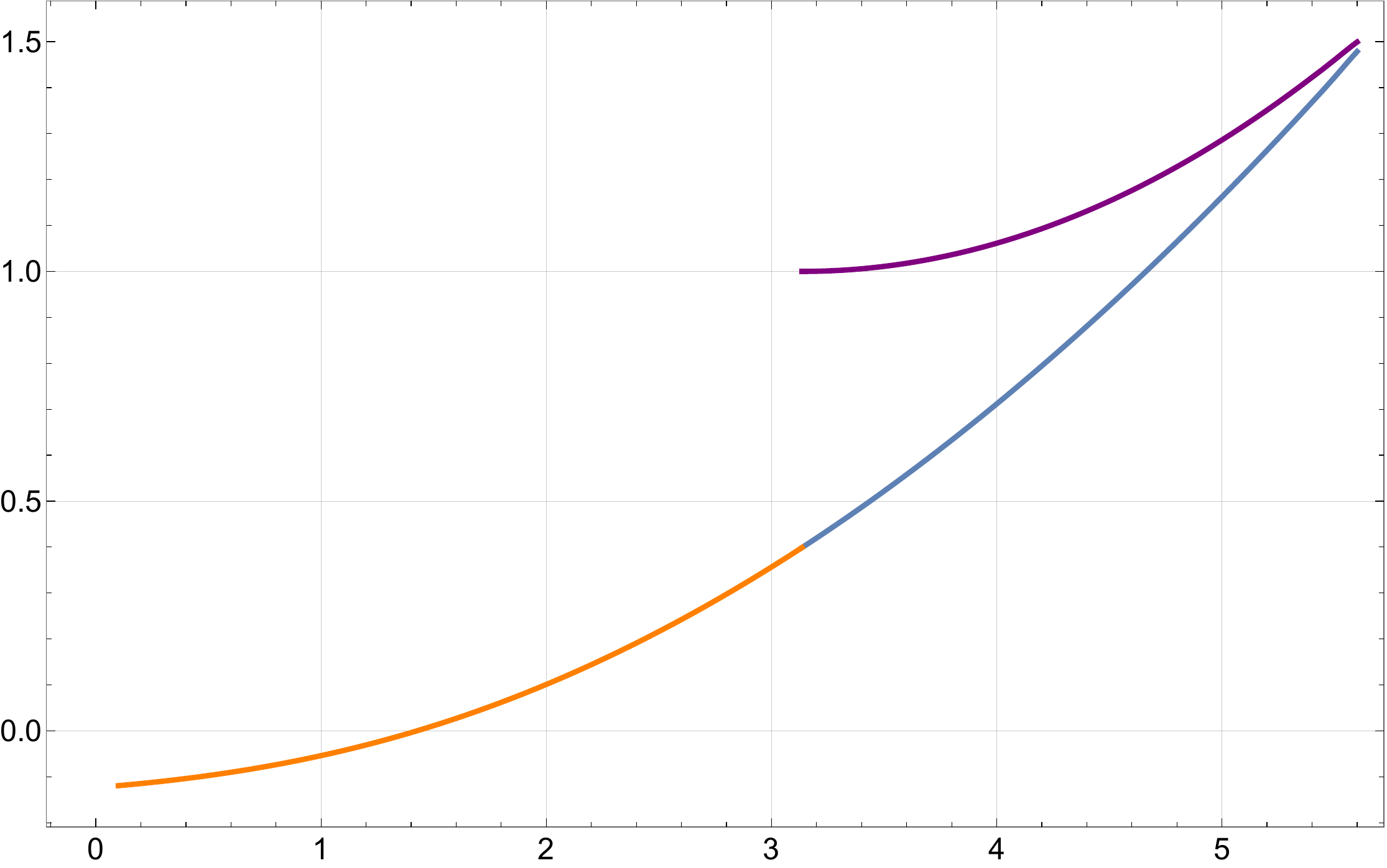}
		\put (31,8) {$\scriptstyle E_{\text{rec.}}/E_{\text{con.}}$}
		\put (74,53) {$\scriptstyle E_{\text{rec.}}^L/E_{\text{con.}}$}
		\put (87,45) {$\scriptstyle E_{\text{rec.}}^S/E_{\text{con.}}$}
		\put (101,-1) {$h/\ell$}
		\put (82.5,3){\tikz \draw[line width=0.35mm,dashed,black] (0,0)--(0,4.4);}
		\put (57.6,3){\tikz \draw[line width=0.35mm,dashed,red] (0,0)--(0,6.4);}
	\end{overpic}
\vspace{0.3cm}
	\caption{Ratio of energies for the different configurations as a function of $h/\ell$, for the values $L =2 \ell$ and $\varphi = \pi/16$. The red dashed line represents the value $\tilde h/\ell = \sqrt{\pi^2 - \varphi^2}$ above which two reconnected configurations exist. The black dashed line represents the value $h^*/\ell$, where $E^{S}_{\text{rec.}}/E_{\text{con.}} = 1$ and the first order phase transition occurs.}\label{Fig: ratios 2}
\end{figure}
\end{itemize}

All in all, we then see that the brane system behaves qualitatively the same, independent of the value of $\alpha$. For the ease of the reader, we summarize below the different cases and the associated phase transitions.

\subsubsection*{Summary}
When $L \leq \ell_\varphi$ and $h <  \tilde h = \ell \sqrt{\pi^2 - \varphi^2}$, there are two brane configurations, a reconnected configuration and a connected one, and the former is always energetically favorable compared to the latter. As the value of $h$ increases and passes $\tilde h$, the two configurations become the same and a second order phase transition occurs at $h=\tilde{h}$. 

When $L >  \ell_\varphi$ and $h < \tilde h$ there is one reconnected configuration that is always energetically favorable with respect to the connected one, as for $L \leq \ell_\varphi$. However, when $h \geq \tilde h$, there are three brane configurations: two reconnected and one connected. The $\theta_L$ reconnected configuration is unstable, having maximal energy. The $\theta_S$ and the connected configurations represent a global and a local minimum, respectively, whenever $h<h^\ast$. For $h>h^\ast$, the role of the two solutions exchange and the connected one becomes an absolute minimum. So, as $h$ increases, the brane configurations undergo a first order phase transition at $h=h^\ast$.

It is worth noting that for small supersymmetry breaking parameter, $\alpha \sim 0$, one gets that $\ell_\varphi \sim \alpha^{-1/2} \rightarrow \infty$ and the range in which the phase transition is second order, {\it i.e.} $L \leq \ell_\varphi$, can be made parametrically large.

%%%%%%%%%%%%%%%%%%%%%%%%%%%%%%
\subsection{On the tachyonic origin of the phase transition}
\label{tachyon}

In section \ref{PTN}, by computing energies of brane webs in the limit of a large number $N$ of $(1,1)$ 5-branes, we have shown that a phase transition of first or second order occurs between a connected and a reconnected configuration, as one varies $h$, at fixed $L$. As in the simplest setup of the $E_1$ theory \cite{Bertolini:2021cew}, the instability of the connected brane web against decay to the reconnected one is expected to originate from a tachyonic mode of an open $(1,-1)$ string stretched between the $(1,-1)$ 5-branes which develops for small enough $h$. 

Let us start considering two D5 branes at an angle $\alpha$. At weak string coupling, the spectrum of the strings ending on the branes can be explicitly calculated and the modes localized at the intersection are tachyonic with mass $m_T^2\sim - 2 \pi \alpha \,\ell_s^2 T_{(1,0)}^2$. This holds both at small angles $\alpha$ and at large angles $\alpha \sim \pi$. Separating the D5 at a distance $h$, the lowest excitations develop an additional positive mass $\sim h^2 T_{(1,0)}^2$ since the minimal length of these strings is now $h$. So, when $h^2=\tilde{h}_{\text{flat}}^2\sim 2 \pi \alpha \ell_s^2$, the lowest mode becomes massless and the system is locally stable. This is expected to remain true also at strong $g_s$ coupling, as was argued in \cite{Giveon:2007fk} in the case of two D4 branes at angles. 

Since this brane system is $SL(2,\mathbb{Z})$ dual to a system of two $(1,-1)$ 5-branes, one can argue that also in this latter system a tachyonic mode is present at small enough distance between the branes, while for $h^2\sim \alpha \ell_s^2$ the configuration should become locally stable.

Our previous analysis shows that this is what actually happens for $L\leq \ell_\varphi$:\footnote{Remind that $\ell_\varphi =\frac{\pi \ell}{\sqrt{\pi^2 - \varphi^2}}$ with $\ell = 2^{1/4 }\ell_s \sqrt{N}$.} there is a phase transition at $h=\tilde{h}$ and, for $h>\tilde{h}$, the connected configuration becomes an absolute minimum of the energy system. At this point, $\tilde{h}\sim \sqrt{\alpha}$ for both $\alpha\sim 0$ and $\alpha\sim \pi$, so we expect the tachyon to condense and to be responsible for the second order phase transition.

For $L >  \ell_\varphi$, the connected configuration ceases to be a maximum at $h\sim \tilde{h}$ but remains globally unstable until $h=h^\ast$. At that point, this is energetically favorable and becomes the absolute minimum of the configuration energy. So at $h\sim \tilde{h}$, the local instability is resolved when the tachyon becomes massless, but a non-perturbative one remains until $h\sim h^\ast$. This realizes the first order phase transition we saw in section \ref{PTN}. 

Note that our transition point $\tilde{h}$ is of order $\sqrt{N}$, while the tachyonic mass between the branes is expected to be $\sim \mathcal{O}(1)$. The same mismatch was found in \cite{Giveon:2007fk} in the case of two D4 branes at an $\alpha=\pi$ angle in a background of $N$ NS5 branes. This apparent tension of the parameters was related to the presence of the NS5 stack\footnote{In their case, the angle was fixed to $\alpha=\pi$.} which was found to modify the tachyonic contribution to the mass as
\begin{equation}
m_T^2\sim - \pi \alpha \,\ell^2 T_{(1,0)}^2. 
\end{equation}
Again, this was argued to remain true also at strong $g_s$ coupling. 

Although the system in \cite{Giveon:2007fk} is only $SL(2, \mathbb{R})$ dual to ours, we find the same behavior for our brane set-up at $\alpha=\pi$ and a similar transition at $\alpha\neq \pi$. We are then led to conclude that also in our case the second order phase transition is mediated by a tachyon  becoming massless at $h\sim \tilde{h}$. Figures \ref{Fig: Potential for rho<l} and \ref{Fig: Potential for rho>l} provide a qualitative behavior of the tachyon potential whose minimum, the tachyon VEV, goes smoothly to zero as $h$ is varied or jumps abruptly when the transition is, respectively, second order, figure \ref{Fig: Potential for rho<l} or first order, figure \ref{Fig: Potential for rho>l}.

%%%%%%%%%%%%%%%%%%%%
%%%%%%%%%%%%%%%%%%%%
\section{Discussion}
\label{discussion}

In this paper we have considered a generalization of the supersymmetry breaking deformation of the $E_1$ theory proposed in \cite{BenettiGenolini:2019zth}, by considering a similar setup for the $X_{1,N}$ theory. The response of the system upon this supersymmetry breaking deformation is qualitatively similar to the $E_1$ case \cite{Bertolini:2021cew}. In particular, considering both the supersymmetry preserving and the supersymmetry breaking deformations at once, it was shown that the parameter space is divided in two different regions separated by a phase transition. For the $E_1$ theory, the order of the phase transition could not be unequivocally established. In the present case, instead, thanks to the possibility of taking $N$ large, it was possible to characterize the phase transition, which, in a certain regime of parameters, was shown to be second order. This gives evidence for the existence of non-supersymmetric fixed points in five dimensions.

One could wonder whether finite $N$ corrections could change this state of affairs. Following arguments similar to those in \cite{Giveon:2007fk}, whose brane system is similar to ours, one could argue that no qualitative difference is expected. Note, however, that while finite $N$ corrections modify both brane systems, an advantage of the system considered in \cite{Giveon:2007fk} is that a small string coupling limit can be taken in which $1/N$ corrections can in principle be computed. This is not the case for our brane web, whose structure changes as the string coupling is modified. 

Another aspect which deserves attention has to do with the dependence of our result on the fixed length $L$ of the 5-brane prongs. In particular, as $L$ crosses $\ell_\varphi$ from the bottom, the phase transition turns from being second order to be first order. For one thing, in the supersymmetric limit $L$ is not a relevant parameter, as the five-dimensional dynamics of the system is independent of $L$ (indeed, one can send the 7-branes on which the 5-brane prongs end all the way to infinity without any change in the dynamics \cite{DeWolfe:1999hj}). This does not seem to be the case after we break supersymmetry. From the 7-brane theory point of view, this does not come as a surprise, since $L$ is related to a Coulomb branch modulus of the eight-dimensional theory living on the 7-branes. By rotating the brane system this modulus is lifted, but only a detailed study of the 7-brane dynamics could tell whether this would be stabilized to some finite value or, say, sent all they way to infinity. This is hard to figure out, since the brane system is intricate and more complicated than a system of branes at angle in isolation. This is an important aspect worth investigate further, even though present string techniques do not seem to be enough to tackle it. This said, it is reassuring that whenever the phase transition is second order, the value of $h$ at which the phase transition occurs, $h=\tilde  h$, does not depend on $L$. Notice, further, that if the supersymmetry breaking deformation is taken to be small, $\ell_\varphi$ can be made parametrically large and hence one can take $L$ large as well, still having the phase transition being second order. In this regime the 7-branes are far from the stack compared to the scale $\tilde h$ at which the transition happens. Therefore, the 7-brane metric, which would change non-trivially the background and which we have not considered in our analysis, would not have any sensible effect on the dynamics triggering the phase transition. 

The property of the $X_{1,N}$ theory may be shared by other systems, some of which could also admit an holographic dual description. While no fully stable non-supersymmetric $AdS_6$ backgrounds are known (see \cite{Bena:2020xxb,Suh:2020rma,Itsios:2021xwh,Apruzzi:2021nle} for recent works addressing this point),  
this is yet an interesting and potentially far reaching direction to be pursued.

We hope to return on some of these issues in a future work.

%========================================
\subsection*{Acknowledgements}
%========================================

We are grateful to Oren Bergman and Diego Rodriguez-Gomez for fruitful exchange of ideas and enlightnening comments, for drawing our attention to ref. \cite{Giveon:2007fk}, and for useful feedbacks on a preliminary draft version. We also thank Riccardo Argurio, Antoine Bourget, Lorenzo Di Pietro, Marco Fazzi, Gabriele Lo Monaco and  Christoph Uhlemann for discussions. This work is partially supported by MIUR PRIN Grant 2020KR4KN2 "String Theory as a bridge between Gauge Theories and Quantum Gravity" and by INFN Iniziativa Specifica ST\&FI. JvM is also supported by the ERC-COG grant NP-QFT No. 864583 "Non-perturbative dynamics of quantum fields: from new deconfined phases of matter to quantum black holes". 

\bibliography{bib5d} 
\bibliographystyle{JHEP} 

\end{document}